\documentclass[aps,twocolumn,showpacs,amsmath,amssymb,prc]{revtex4-2}
\usepackage{graphicx,here}
\usepackage{multirow}
\usepackage{tikz}
\usepackage{epstopdf}
\usepackage[percent]{overpic}
\usepackage{scrextend}
\usepackage{xcolor,xspace}
\usepackage[ulem=normalem]{changes}
\usepackage{hyperref}
\hypersetup{colorlinks=True,urlcolor=blue,linkcolor=blue,citecolor=blue,filecolor=black}

\colorlet{Changes@Color}{red}
\usepackage{dcolumn,color,footnote,bm,braket}
\usepackage{url,longtable,tabularx}
\usepackage{threeparttable}

\usepackage{fancybox}
\usetikzlibrary{matrix}

\newcommand\+{\dagger}
\newcommand\jr{j_{\rho}}

\newcommand\mzn{M^{(0\nu)}}

\newcommand\mgt{M_{\mathrm{GT}}^{(0\nu)}}
\newcommand\mfe{M_{\mathrm{F}}^{(0\nu)}}
\newcommand\mte{M_{\mathrm{T}}^{(0\nu)}}

\newcommand\hb{\hat{H}_{\mathrm{B}}}

\newcommand\db{\beta\beta}
\newcommand\tnbb{2\nu\beta\beta}
\newcommand\znbb{0\nu\beta\beta}
\newcommand\ga{g_{\mathrm{A}}}
\newcommand\gv{g_{\mathrm{V}}}

\newcommand\taubb{T_{1/2}^{(0\nu\beta\beta)}}

\begin{document}

\title{Neutrinoless $\beta\beta$ decay in the interacting boson model based on the nuclear energy density functionals}

\author{Kosuke Nomura}
\email{nomura@sci.hokudai.ac.jp}
\affiliation{Department of Physics, 
Hokkaido University, Sapporo 060-0810, Japan}
\affiliation{Nuclear Reaction Data Center, 
Hokkaido University, Sapporo 060-0810, Japan}

\date{\today}

\begin{abstract}
The neutrinoless $\beta\beta$
($0\nu\beta\beta$) decay
nuclear matrix elements (NMEs)
are calculated in the interacting boson model (IBM),
which is based on the nuclear
energy density functional (EDF) theory.
The Hamiltonian of the IBM that gives rise to 
the energies and wave functions of the ground 
and excited states of $0\nu\beta\beta$ decay 
emitting isotopes and corresponding final nuclei 
is determined by mapping the self-consistent 
mean-field deformation-energy surface obtained with 
a given EDF onto the corresponding bosonic energy surface. 
The transition operators are formulated 
using the generalized seniority scheme, 
and the pair structure 
constants are determined by the inputs 
provided by the self-consistent calculations.
The predicted values of the $0\nu\beta\beta$-decay 
NMEs with the nonrelativistic and 
relativistic EDFs are compared with those
resulting from different many-body methods. 
Sensitivities of the predicted NMEs to the 
model parameters and assumptions
are discussed. 
\end{abstract}

\maketitle

\section{Introduction}

Nuclear $\beta\beta$ decay is a rare process 
for the even-even nucleus with mass $A$ 
and proton $Z$ numbers $(A,Z)$
to decay into the one 
with $(A,Z\pm2)$,
emitting two electrons or positrons 
\cite{mayer1935}.
Two-neutrino $\beta\beta$ ($\tnbb$) decays, 
which are accompanied by the emissions of 
two antineutrinos or neutrinos, are allowed 
transitions in the standard model of elementary 
particles and have been observed experimentally. 
Another type of the $\db$ decay
that may exist is the neutrinoless
$\beta\beta$ ($\znbb$) decay,
which does not emit neutrinos.
The search for the $\znbb$ decay
is of fundamental importance,
since its observation
would imply some new physics
beyond the standard model because
this decay process violates 
symmetry requirements 
of the electroweak interaction such as 
the lepton-number conservation law,
and provide a crucial piece of information 
as to the mass and nature of the neutrino,
i.e., if it is a Dirac or Majorana particle
\cite{majorana1937}.
Experiments that are aimed to detect the 
$\znbb$ decay have been operational 
around the world, and next-generation experiments 
are also prepared (see recent reviews e.g., 
Refs.~\cite{avignone2008,agostini2023,gomezcadenas2024}).

Theoretical predictions have been made for
the $\znbb$-decay nuclear matrix elements (NMEs) by 
various nuclear many-body approaches,
including the quasiparticle
random-phase approximation (QRPA)
\cite{mustonen2013,hyvarinen2015,simkovic2018,fang2018,terasaki2020,jokiniemi2023},
nuclear shell model (NSM)
\cite{horoi2016,iwata2016,menendez2018,corragio2020,tsunoda2023,jokiniemi2023,castillo2025},
neutron-proton interacting boson model (IBM-2)
\cite{barea2009,barea2013,barea2015,deppisch2020},
the generator coordinate method (GCM) using the
energy density functional (EDF)
\cite{trodriguez2010,vaquero2013,song2017}, 
and {\it ab initio} methods
including the
in-medium similarity renormalization group (IMSRG)
approach \cite{yao2020,belley2021,belley2024},
coupled cluster (CC) theory \cite{novario2021},
and effective field theory (EFT) \cite{brase2022}.
These studies were
reported for the last decades,
and more comprehensive lists of the
relevant theoretical studies
are found in recent review articles
\cite{avignone2008,agostini2023,gomezcadenas2024,engel2017}.
Predicted $\znbb$ NMEs resulting
from these
many-body methods differ by several factors.
To reduce these differences, it is important to
improve accuracy and identify uncertainties of
a given nuclear structure theory.

In particular, the IBM, a model 
in which correlated pairs of valence nucleons 
are represented by bosons,
has been employed for 
investigations of the $\znbb$ and $\tnbb$ 
decays \cite{barea2009,barea2013,barea2015,deppisch2020}. 
In these studies, the Gamow-Teller (GT), Fermi, 
and tensor transition operators were derived by 
using a fermion-to-boson mapping that 
is based on the generalized seniority scheme 
of the nuclear shell model 
\cite{OAIT,OAI}.
The nuclear wave functions for the even-even 
parent and daughter nuclei were computed by 
the IBM Hamiltonian with the parameters 
directly fitted to reproduce the experimental 
low-energy spectra.

In the present study, $\znbb$-decay 
NME predictions are made within the IBM 
that is based on the nuclear EDF theory  
\cite{nomura2008,nomura2010}. 
In this framework,
the Hamiltonian of the IBM that produces 
energies and wave functions of even-even nuclei 
is determined by mapping the potential 
energy surface (PES) obtained from 
the self-consistent mean-field (SCMF) calculations 
based on a given EDF onto the corresponding 
energy surface in the boson system.
The Hamiltonian parameters are
completely determined
by using the inputs obtained from
the SCMF calculations,
for which any phenomenological
adjustment to experiment as in the conventional
IBM studies is not needed.
This allows one
to predict low-lying states
of those nuclei that are far from the stability
and for which experimental data do not exist.
This work adopts
the method of Ref.~\cite{barea2009},
in which the GT, Fermi,
and tensor transition operators
are formulated
in terms of the generalized seniority.
However, pair structure constants
that are included in the 
coefficients for the transition operators  
are computed by using the results of 
the SCMF calculations.

The IBM mapping procedure
has recently been applied to
the calculations for the
$\tnbb$-decay NMEs for 13 candidate 
even-even nuclei \cite{nomura2022bb,nomura2024bb}, 
in which states of the 
intermediate odd-odd nuclei were
explicitly calculated in terms of the 
interacting boson-fermion-fermion model (IBFFM) \cite{IBFM}. 
In Refs.~\cite{nomura2022bb,nomura2024bb}, 
the even-even core (or IBM) Hamiltonian 
was determined by the mapping procedure, 
and the single-particle energies and 
occupation probabilities, which are essential building 
blocks of the IBFFM Hamiltonian, GT and Fermi transition operators, 
were also provided by the SCMF calculations based 
on a relativistic EDF. 
Remaining coupling constants of the boson-fermion
and odd neutron-proton interactions were, however, 
fitted to reproduce low-energy levels of odd-odd nuclei.

The present study exploits many of the 
model ingredients from 
Refs.~\cite{nomura2022bb,nomura2024bb}, including 
the SCMF PESs obtained with the relativistic EDF, 
and derived IBM parameters. 
As in the earlier IBM and in the majority of other
theoretical calculations for the $\znbb$ decay, 
the present study assumes closure approximation, 
that is, intermediate states of 
the neighboring odd-odd nuclei are neglected 
and their energies are represented by some average energy. 
This approximation is justified for the 
$\znbb$ decay, since the neutrino momenta 
in this process are of the order of 100 MeV, which 
is far above the typical nuclear excitation energies.
The present framework thus consists of 
the calculations on the even-even parent 
and daughter nuclei, 
and the $\znbb$ transition matrix elements.
The NMEs are calculated for the 
proposed $\znbb$ emitters $^{48}$Ca, 
$^{76}$Ge, $^{82}$Se, $^{96}$Zr, $^{100}$Mo, 
$^{116}$Cd, $^{128}$Te, $^{130}$Te, $^{136}$Xe, 
and $^{150}$Nd, and are compared with other
theoretical predictions.
To show the robustness of the
IBM mapping,
in addition to the relativistic
SCMF calculations, 
nonrelativistic calculations employing 
the Gogny interaction are here performed. 
Sensitivities of the NME predictions 
to several model parameters and assumptions 
are studied, including the   
choice of the EDFs, IBM Hamiltonian parameters,
SCMF-to-IBM mapping procedure, and pair structure constants
for the transition operators.

Section~\ref{sec:theory} gives 
a brief description of the IBM mapping
procedure, and the formalism for
calculating the $\znbb$-decay NMEs.
The intrinsic and low-energy spectroscopic 
properties of even-even nuclei involved in the 
$\znbb$ decays are discussed in Sec.~\ref{sec:str}. 
Section~\ref{sec:db} presents the predicted 
$\znbb$ NMEs and half-lives. 
The sensitivity analyses of the NME predictions
are given in Sec.~\ref{sec:model}. 
A summary of the principal results and 
perspectives for possible extensions of the 
model are given in Sec.~\ref{sec:summary}.

\section{Method to calculate $\znbb$-decay
nuclear matrix elements\label{sec:theory}}

\subsection{IBM-2 mapping procedure}

Microscopic inputs to the IBM are results of 
the constrained SCMF calculations performed 
for even-even nuclei that are parents
and daughters for the $\znbb$ decays
$^{48}$Ca $\to$ $^{48}$Ti, 
$^{76}$Ge $\to$ $^{76}$Se, 
$^{82}$Se $\to$ $^{82}$Kr, 
$^{96}$Zr $\to$ $^{96}$Mo, 
$^{100}$Mo $\to$ $^{100}$Ru, 
$^{116}$Cd $\to$ $^{116}$Sn, 
$^{128}$Te $\to$ $^{128}$Xe, 
$^{130}$Te $\to$ $^{130}$Xe, 
$^{136}$Xe $\to$ $^{136}$Ba, and 
$^{150}$Nd $\to$ $^{150}$Sm.
Two EDFs are considered in the present study:
the density-dependent point-coupling (DD-PC1) 
interaction \cite{DDPC1},
a representative set of parameters
for the relativistic EDF, 
and the Gogny D1M \cite{D1M} interaction 
as a representative of the nonrelativistic
energy functionals.
The SCMF calculations are carried out using 
the framework of the relativistic Hartree-Bogoliubov (RHB)
method \cite{vretenar2005,niksic2011,DIRHB,DIRHBspeedup},
and the Hartree-Fock-Bogoliubov
(HFB) method \cite{robledo2019}
for the nonrelativistic Gogny interaction. 
In the RHB-SCMF calculations,
a separable pairing force of finite range 
\cite{tian2009} is considered.
The constraints imposed on the self-consistent 
calculations are on the mass quadrupole
moments $Q_{20}$ and $Q_{22}$, that is,
the neutron and proton quadrupole
moments are calculated separately,
but the sums should be made equal to the
desired values of $Q_{20}$ and $Q_{22}$.
The quadrupole moments $Q_{20}$ and $Q_{22}$
are related to the polar deformation variables
$\beta$ and $\gamma$, describing the
triaxial quadrupole shapes \cite{BM}.
A set of the constrained RHB or HFB
self-consistent calculations yields 
the PES in terms of the $(\beta,\gamma)$ deformations, 
$E_{\mathrm{SCMF}}(\beta,\gamma)$, which is 
to be used to construct the IBM Hamiltonian.
Note that, in the following, the calculations with microscopic
inputs from the constrained RHB method with
the DD-PC1 EDF and the constrained
HFB method with the Gogny-D1M EDF
are referred to as RHB and HFB,
respectively, to distinguish the relativistic
from the nonrelativistic SCMF frameworks.

The present study employs 
the neutron-proton IBM (IBM-2)
\cite{OAIT,OAI,IBM}, 
which comprises neutron
$s_{\nu}$ and $d_{\nu}$ bosons, and 
proton $s_{\pi}$ and $d_{\pi}$ bosons. 
$s_{\nu}$ ($s_{\pi}$) and $d_{\nu}$ ($d_{\pi}$) bosons 
represent collective monopole and quadrupole pairs 
of valence neutrons (protons), respectively. 
The number of neutron (proton) bosons, denoted by 
$N_\nu$ ($N_{\pi}$), is equal to 
the number of valence neutron
(proton) pairs
and is counted from the nearest
neutron (proton) closed shell.
In the present cases,
the neutron and proton closed
shells are taken to be
$(N,Z)=(28,20)$
for those nuclei with mass $A=48$,
$(N,Z)=(28,50)$
for the $A=76$ and 82 nuclei,
$(N,Z)=(50,50)$ for
the $A=96$ and 100 nuclei,
and $(N,Z)=(82,50)$ for
the $A=116$, 128, 130, 136,
and 150 nuclei.

The IBM-2 Hamiltonian adopted in this work
takes the form
\begin{align}
\label{eq:hb}
 \hb =\,
\epsilon_{d}(\hat{n}_{d_{\nu}}+\hat{n}_{d_{\pi}})
+\kappa\hat{Q}_{\nu}\cdot\hat{Q}_{\pi}
+ \hat V_{\rm cub} \; .
\end{align}
$\hat{n}_{d_\rho}=d^\+_\rho\cdot\tilde d_{\rho}$ 
($\rho=\nu,\pi$)
is the number operator of $d$ bosons,
with $\epsilon_{d}$ being the single $d$-boson
energy relative to the $s$-boson energy, and 
$\tilde d_{\rho\mu}=(-1)^\mu d_{\rho-\mu}$. 
The second term is the quadrupole-quadrupole 
interaction between neutron and proton bosons, 
with $\kappa$ being the strength parameter, 
and with 
$\hat Q_{\rho}=d_{\rho}^\+ s_{\rho} + s_{\rho}^\+\tilde d_{\rho} + \chi_{\rho}(d^\+_{\rho}\times\tilde{d}_{\rho})^{(2)}$ being 
the quadrupole operator in the boson system. 
$\chi_\nu$ and $\chi_\pi$ are dimensionless parameters, 
and determine whether the nucleus is 
prolate or oblate deformed. 
The last term represents a cubic or three-body 
term of the form
\begin{align}
\label{eq:cub}
 \hat V_{\rm cub} = \sum_{\rho\neq\rho'}
\theta_{\rho}
[d^\+_{\rho} \times d^\+_{\rho} \times d^\+_{\rho'}]^{(3)}
\cdot
[\tilde d_{\rho'} \times \tilde d_{\rho} \times \tilde d_{\rho}]^{(3)} \; ,
\end{align}
where the strength parameters $\theta_\nu$ 
for neutrons 
and $\theta_{\pi}$ for protons are assumed to be equal, 
$\theta_\nu = \theta_{\pi} \equiv \theta$. 
The cubic term is specifically required 
in order to produce a triaxial minimum in 
the energy surface for $\gamma$-soft nuclei. 
Effects of the cubic term are to lower the $2^+$
bandhead and other members of the
$\gamma$-vibrational band, and to improve the description
of its energy-level systematic and $B(E2)$ transition
strengths. This term, however, does not
affect properties of states in the ground-state band
\cite{nomura2012tri}.

In those nuclei corresponding to $N_{\pi}=0$ 
or/and $N_{\nu}=0$, the 
unlike-boson quadrupole-quadrupole interaction 
$\hat{Q}_{\nu}\cdot\hat{Q}_{\pi}$ 
in (\ref{eq:hb}) vanishes. 
For the semimagic nuclei $^{116}$Sn and $^{136}$Xe, 
in particular, which have $N_{\pi}=0$ 
and $N_{\nu}=0$, respectively, 
a Hamiltonian of the following form is considered:
\begin{align}
\label{eq:hb-semimagic}
 \hb = 
\epsilon_{d_\rho} \hat{n}_{d_{\rho}} 
+\kappa_\rho \hat{Q}_{\rho}\cdot\hat{Q}_{\rho} \; ,
\end{align}
which consists only of the interaction 
terms between like bosons. 
As regards the doubly magic nucleus $^{48}$Ca,
for which $N_{\nu}=N_{\pi}=0$, any IBM Hamiltonian 
does not produce an energy spectrum.

The bosonic energy surface 
$E_{\mathrm{IBM}}(\beta,\gamma)$ is 
calculated as an expectation 
value of the IBM-2 Hamiltonian (\ref{eq:hb}), 
i.e., 
$E_{\mathrm{IBM}}(\beta,\gamma)
=\braket{\Phi|\hb|\Phi}/\braket{\Phi|\Phi}$. 
Here $\ket{\Phi}$ denotes a boson
coherent state, which is defined as
\cite{dieperink1980,ginocchio1980,bohr1980} 
\begin{align}
\label{eq:coherent}
 \ket{\Phi}=
\frac{1}{\sqrt{N_{\nu}!N_{\pi}!}}
(\lambda_\nu^\+)^{N_{\nu}}
(\lambda_\pi^\+)^{N_{\pi}}
\ket{0} \; ,
\end{align}
with
\begin{align}
\label{eq:coherent-1}
 \lambda^{\+}_{\rho}=
s^\+_{\rho}
+\beta_{\rho}\cos{\gamma_{\rho}} d^\+_{\rho 0}
+\frac{1}{\sqrt{2}}\beta_{\rho}\sin{\gamma_{\rho}} 
(d^\+_{\rho 2}+d^\+_{\rho -2}) \; .
\end{align}
The state $\ket{0}$ in (\ref{eq:coherent}) 
represents the boson vacuum, i.e., the inert core. 
In (\ref{eq:coherent-1}), 
$\beta_{\rho}$ and $\gamma_{\rho}$ are amplitudes 
that are considered to be 
boson analogs of the $(\beta,\gamma)$
deformations in the geometrical model.
The neutron $\beta_{\nu}$
and $\gamma_{\nu}$ deformations
are assumed to be equal to those for
protons, $\beta_{\pi}$
and $\gamma_{\pi}$, respectively,
$\beta_{\nu}=\beta_{\pi}\equiv\bar\beta$ 
and $\gamma_{\nu}=\gamma_{\pi}\equiv\bar\gamma$.
These assumptions are made
in order to associate
the PES of the IBM-2 with that of the SCMF.
In the SCMF framework, while the neutron and proton degrees of
freedom are distinguished, the PES is obtained
as a function of the $(\beta,\gamma)$ deformations
as the constraints to mass quadrupole moments are imposed.
Furthermore, the bosonic $\beta$
deformation is assumed to be proportional
to the fermionic counterpart,
$\bar\beta = C_{\beta}\beta$ with $C_{\beta}$ being 
a constant of proportionality,
while the bosonic $\gamma$
deformation is assumed to be identical
to the fermionic one, $\bar\gamma = \gamma$ 
\cite{ginocchio1980,nomura2008}.

The parameters for the Hamiltonian \eqref{eq:hb}
[or \eqref{eq:hb-semimagic}] are determined by
the SCMF-to-IBM mapping \cite{nomura2008,nomura2010}
so that the approximate equality
\begin{align}
\label{eq:map}
 E_{\mathrm{SCMF}}(\beta,\gamma)
\approx
E_{\mathrm{IBM}}(\beta,\gamma)
\end{align}
should be satisfied 
in the vicinity of the global mean-field minimum.
The optimal IBM-2 parameters 
are obtained so that basic characteristics of the 
SCMF PES in the neighborhood of the 
global mean-field minimum, e.g., curvatures 
in the $\beta$ and $\gamma$ deformations, 
and depth and location of the minimum, 
should be reproduced by the IBM-2 PES.
Details of the mapping procedure 
are found in Refs.~\cite{nomura2008,nomura2010}.

The parameters of the IBM-2 Hamiltonian determined 
by the mapping are summarized in 
Tables~\ref{tab:para-ddpc} and \ref{tab:para-d1m} 
in Sec.~\ref{sec:para}.
The mapped IBM-2 Hamiltonian is diagonalized in the 
boson $m$-scheme basis \cite{nomura2012tri,nomura2012phd},
giving rise to the energies and 
wave functions of the ground and excited states 
of the even-even nuclei that are
parents and daughters of the $\znbb$ decays.
By the diagonalization of the mapped IBM-2 Hamiltonian
in the laboratory frame,
some essential correlations that are absent
in the static mean-field approximations
are properly taken into account, such as those
dynamical correlations related to the restoration
of broken symmetries, and quantum fluctuations near
the mean-field minimum.

\subsection{$\znbb$-decay NME\label{sec:znbb}}

The following discussion focuses on the 
simplest case of $\znbb$ decay, that is, 
only the light neutrino exchange and 
long-range part of the NME are considered. 
The half-life of the $\znbb$ decay is 
expressed as
\begin{eqnarray}
\label{eq:tau}
 \left[\taubb\right]^{-1}
= G_{0\nu} \ga^4 |\mzn|^2 
\left(
\frac{\braket{m_\nu}}{m_e}
\right)^2 \; ,
\end{eqnarray}
where $G_{0\nu}$, $\ga$, $\mzn$, 
$\braket{m_\nu}$, and $m_e$ are phase-space factor 
for the $\znbb$ decay, axial-vector coupling constant, 
NME, average light neutrino mass, 
and electron mass, respectively. 
$\mzn$ consists of 
the Gamow-Teller (GT), Fermi (F),
and tensor (T) components:
\begin{eqnarray}
 \mzn = \mgt - \left(\frac{\gv}{\ga}\right)^2 \mfe + \mte \; ,
\end{eqnarray}
where $\gv$ is the vector coupling constant. 
The $\gv$ and $\ga$ values are taken to be 
$\gv=1$ and $\ga=1.269$ \cite{Yao2006}, respectively. 
The matrix elements of the components in $\mzn$ 
for the $0^+ \to 0^+$ $\znbb$ decay are 
computed by using the wave functions of the 
initial $\ket{0^+_i}$ state in parent 
and final $\ket{0^+_f}$ state in daughter nuclei:
\begin{eqnarray}
\label{eq:mzn-0}
 M_\alpha^{(0\nu)} = 
\braket{0^+_f| \hat O_\alpha |0^+_i} \; .
\end{eqnarray}
Here
 \begin{align}
\label{eq:oalph}
 \hat O_\alpha = 
&\frac{1}{2} A_\alpha
\sqrt{\frac{4\pi}{2\lambda+1}}
\sum_{i,j}
\tau^\+_i \tau^\+_j
\nonumber\\
&\times 
H_\alpha(r_{ij})
Y^{(\lambda)}(\Omega_{ij})\cdot
\left[
\Sigma_i^{(s_1)}
\times
\Sigma_j^{(s_2)}
\right]^{(\lambda)}
\end{align}
denotes the corresponding operator, 
and $\alpha=$ F, GT, or T represents 
a set of quantum numbers 
$\lambda$, $s_1$, and $s_2$, 
which specifies the type of the transition: 
$\alpha=$ F for $\lambda=0$ and $s_1=s_2=0$, 
$\alpha=$ GT for $\lambda=0$ and $s_1=s_2=1$, and 
$\alpha=$ T for $\lambda=2$ and $s_1=s_2=1$. 
The factor $A_\alpha$ equals 1, $-\sqrt{3}$, 
and $\sqrt{2/3}$ for Fermi, GT, and tensor transitions, 
respectively. 
The spin operator 
$\Sigma^{(0)}=1$ and $\Sigma^{(1)}=\boldsymbol{\sigma}$,
and $\tau^\+$ stands for the isospin raising operator. 
$Y^{(\lambda)}$ is the spherical harmonics of rank $\lambda$. 
$H_\alpha(r_{ij})$ stands for the radial part of the
neutrino potential
[denoted $V(r)$ in Ref.~\cite{barea2009}]
and, in momentum representation,
is expressed as
\begin{align}
 H_\alpha(r) 
= \frac{2R}{\ga^2}\int 
h^\alpha (p) j_\lambda (pr) p^2 dp \; ,
\end{align}
where $j_\lambda$ is the spherical Bessel function
of rank $\lambda$, and the multiplication
by the factor $2R$, with
$R=1.2A^{1/3}$, is to make the NME dimensionless.
The factors $h^\alpha(p)$ for different transition types
are given by \cite{simkovic1999}
\begin{align}
\label{eq:fehoc}
& h^{\rm F}(p) = h^{\rm F}_{VV}(p) \\
\label{eq:gthoc}
& h^{\rm GT}(p) = 
h^{\rm GT}_{AA}(p) + h^{\rm GT}_{AP}(p) +
h^{\rm GT}_{PP}(p) + h^{\rm GT}_{MM}(p) \\
\label{eq:tehoc}
& h^{\rm T}(p) 
= h^{\rm T}_{AP}(p) 
+ h^{\rm T}_{PP}(p) + h^{\rm T}_{MM}(p) \; ,
\end{align}
where the subscripts
$VV$ and $AA$ denote the vector and axial-vector 
couplings, respectively, and the terms indicated 
by the subscripts $PP$, $AP$, and $MM$ represent 
higher-order contributions \cite{simkovic1999}
from pseudoscalar, axial-vector-pseudoscalar, 
and magnetic couplings, respectively.
The factors $h^{\alpha}(p)$ are further expressed
in a product form: 
\begin{align}
 h^{\alpha}(p) = v(p) \tilde h^{\alpha}(p) \; ,
\end{align} 
where $v(p)$ stands for the neutrino potential
\begin{align}
 v(p) = \frac{2}{\pi} \frac{1}{p(p+\tilde A)}  \; .
\end{align}
$\tilde A$ is the closure energy, and its values 
are taken from Ref.~\cite{tomoda1991}.
The exact forms of the form factors $\tilde h^{\alpha}(p)$
are summarized in Sec.~\ref{sec:hoc}.

The operator 
$\hat O_\alpha$ in (\ref{eq:oalph}) is 
rewritten in a second-quantized form,  
\begin{align}
\label{eq:oalph-1}
 \hat O_\alpha
= &- \frac{1}{4}
\sum_{j_1j_2}
\sum_{j_1'j_2'}
\sum_{J}
(-1)^J
\nonumber\\
&\sqrt{1+(-1)^J\delta_{j_1j_2}}
\sqrt{1+(-1)^J\delta_{j_1'j_2'}}
\nonumber \\
&
O_{\alpha}(j_1j_2j_1'j_2';J)
(c^\+_{j_{1}} \times c^\+_{j_{2}})^{(J)}
\cdot
(\tilde c_{j_{1}'} \times \tilde c_{j_{2}'})^{(J)} \; ,
\end{align} 
where
$O_{\alpha}(j_1j_2j_1'j_2';J)$
is the corresponding fermion two-body 
matrix element in the 
two-particle basis $\ket{j_1j_2;JM}$
defined by
\begin{align}
 \ket{j_1j_2;JM} = 
\frac{1}{\sqrt{1+(-1)^J\delta_{j_1j_2}}}
(c^\+_{j_1} \times c^\+_{j_2})^{(J)}_{M} \ket{0} \; ,
\end{align}
where $j_i$ ($i=1,2$) represents
a set of single-particle quantum numbers 
$j_i \equiv \{ n_i,l_i,j_i,m_i \}$, 
and the primed $j_i'$ (unprimed $j_i$) symbol
denotes the neutron (proton) state.
The expression for $O_{\alpha}(j_1j_2j_1'j_2';J)$
is given in Sec.~\ref{sec:tbme}. 
Note that the quantities here denoted
$\hat O_{\alpha}$ \eqref{eq:oalph}
and $O_{\alpha}(j_1j_2j_1'j_2';J)$ \eqref{eq:oalph-1}
correspond to $V_{s_1,s_2}^{(\lambda)}$
and $V_{s_1,s_2}^{(\lambda)}(j_1j_2j_1'j_2';J)$
in Ref.~\cite{barea2009}, respectively.

In addition, the short-range correlation (SRC)
is taken into account by multiplying
$H_\alpha(r_{ij})$ by 
the following Jastrow function squared:
\begin{align}
 f(r) = 1 - c e^{ar^2} (1-br^2) \; ,
\end{align}
with the Argonne parametrization for the $NN$ force, 
$a=1.59$ fm$^{-2}$, 
$b=1.45$ fm$^{-2}$, and $c=0.92$ \cite{simkovic2009}. 
The SRC is incorporated by 
the Fourier-Bessel transform of $f(r)$, 
since the present formulation is in momentum space.

The nuclear many-body calculations are 
required to obtain the matrix element 
\begin{eqnarray}
\label{eq:tnt}
 \braket{0^+_f|
(c^\+_{j_{1}} \times c^\+_{j_{2}})^{(J)}
\cdot
(\tilde c_{j_{1}'} \times \tilde c_{j_{2}'})^{(J)} 
|0^+_i} \; ,
\end{eqnarray}
which appears in $\mzn_\alpha$ (\ref{eq:mzn-0}). 
Here the truncated Hilbert space
consisting of the $S$ ($J=0^+$) and $D$ ($J=2^+$)
collective isovector pairs is considered for neutrons
and protons and the corresponding pair creation
operators are given by
\begin{align}
\label{eq:spair}
& S^\+ = \sum_{j} \alpha_j 
\sqrt{\frac{\Omega_j}{2}}
(c^\+_{j} \times c^\+_{j})^{(0)}
\\
\label{eq:dpair}
& D^\+ = \sum_{j_1j_2} \beta_{j_1j_2} 
\frac{1}{\sqrt{1+\delta_{j_1j_2}}}
(c^\+_{j_1} \times c^\+_{j_2})^{(2)} \; ,
\end{align}
where $\alpha_{j}$ and $\beta_{j_{1}j_{2}}$
denote pair structure
constants.
$\alpha_{j}$ is assumed to be proportional
to the occupation amplitude $v_{j}$
of the neutron or proton in the orbit $j$
in a given nucleus, $\alpha_{j}=Kv_{j}$,
which is provided by the SCMF
calculation performed for the
neighboring odd-odd nucleus with the
constraint to zero deformation
using the
procedure developed in Ref.~\cite{nomura2016odd}.
For the calculation for odd-odd nuclei,
a standard HFB or
RHB method without blocking is used,
but imposing the odd nucleon
number constraint.
The proportionality constant $K$
is obtained by imposing that the
normalization of the $\alpha_{j}$'s
is equal to the maximum number of pairs,
denoted $\Omega$, which
the model space can accommodate
\begin{eqnarray}
\sum_{j}\alpha_{j}^{2}\Omega_{j}=\sum_{j}\Omega_{j}=\Omega \; ,
\end{eqnarray}
where $\Omega_{j}=\left(2j+1\right)/2$
and the sum is taken over all
the single-particle states
in the model space.
Then $\alpha_{j}$ is
computed using the relation
\begin{eqnarray}
\label{eq:alpha}
\alpha_{j}=\sqrt{\frac{\Omega}{\sum_{k}v_{k}^{2}\Omega_{k}}}v_{j} \; . 
\end{eqnarray}
For like-hole configurations ($v_{j}^{2}>0.5$),
$v_{j}$ is replaced with
the unoccupation amplitude,
$u_{j}=(1-v_{j}^{2})^{1/2}$.
The $\alpha_{j}$ values thus obtained
are used to calculate the
coefficients $\beta_{j_{1}j_{2}}$ by
\begin{eqnarray}
\label{eq:beta} 
\beta_{j_{1}j_{2}}=\frac{\alpha_{j_{1}}+\alpha_{j_{2}}}
  {\sqrt{5}\Omega\left(1+\delta_{j_{1}j_{2}}\right)}
  \left\langle j_{1}\left\Vert r^{2}Y^{\left(2\right)}\right\Vert j_{2}
  \right\rangle \; .
\end{eqnarray}
This formula was derived in the
microscopic study of the IBM for
nondegenerate orbits \cite{pittel1982},
in which the $D$-pair operator was
expressed as a commutator
of the $S$-pair and quadrupole operators.
The sign of $\alpha_j$ relative to
that of $\beta_{j_1j_2}$ is assumed
to be consistent with 
that considered in the previous
IBM-2 calculations for the $\znbb$ decay
\cite{barea2009}
by introducing
additional factors $(-1)^l$ for $\alpha_j$ 
and $(-1)^{(l_1-l_2)/2}(-1)^{j_1-j_2+1}$ 
for $\beta_{j_1j_2}$, and changing
the sign of $\alpha_j$
for like-hole configurations.

The single-particle configurations
and $v_j^2$ values
calculated by the SCMF methods for the corresponding
orbits are summarized
in Tables~\ref{tab:vv_48-82}, \ref{tab:vv_96-116},
\ref{tab:vv_128-136}, and \ref{tab:vv_150}
in Appendix~\ref{sec:spe}.
As is customary
in the interacting boson-fermion model calculations
for odd-mass and odd-odd nuclei \cite{IBFM},
in which an explicit coupling of fermionic
to bosonic degrees of freedom should be taken into account,
and in the previous IBM-2 studies
on $\db$ decays \cite{barea2009,barea2013,barea2015},
the single-particle space considered in the present
study corresponds to the same valence space
as that to which the IBM-2 is mapped.
These choices of the single-particle spaces
are justified, on the basis of the fact that
in the present framework
the only microscopic inputs from the SCMF
calculations that concern the single-particle
properties are occupation probabilities $v_j^2$
for given orbits, and only those $v^2_j$ values
for the orbits near the Fermi surfaces
are most relevant, which
indeed correspond to the IBM-2
configuration spaces.
Those single-particle states
that are far from the Fermi surfaces
play only negligible roles,
since they are either fully
occupied ($v_j^2\approx 1$)
or unoccupied ($v_j^2\approx 0$) states,
in which cases
the $\alpha_{j}$ coefficients vanish
in the formula \eqref{eq:alpha}.

As in Ref.~\cite{barea2009}, the shell-model 
$S_{\rho}$ and $D_{\rho}$ 
pairs are mapped onto the $s_{\rho}$ 
and $d_{\rho}$ bosons, respectively. 
The following mapping is considered 
in Eq.~(\ref{eq:tnt}):
\begin{align}
\label{eq:ajr-0}
 & (c^\+_{j} \times c^\+_{j})^{(0)}
\mapsto A_\pi(j) s^\+_\pi
\\
\label{eq:bjr-0}
 & (c^\+_{j_1} \times c^\+_{j_2})^{(2)}
\mapsto B_\pi(j_1j_2) d^\+_\pi
\end{align}
for protons, and 
\begin{align}
\label{eq:ajr-1}
 & (\tilde c_{j'} \times \tilde c_{j'})^{(0)}
\mapsto A_\nu(j') \tilde s_{\nu}
\\
\label{eq:bjr-1}
 & (\tilde c_{j_1'} \times \tilde c_{j_2'})^{(2)}
\mapsto B_{\nu}(j_1'j_2') \tilde d_{\nu}
\end{align}
for neutrons. 
The boson image of $\mzn_\alpha$ therefore reads
\begin{align}
\label{eq:mzn-1}
 \mzn_\alpha
= 
&- \frac{1}{2}
\sum_{j}
\sum_{j'}
O_{\alpha}(jjj'j';0)
\nonumber\\
&\quad
\times A_\pi(j)A_{\nu}(j')
\braket{0^+_f| 
s^\+_\pi \cdot \tilde s_{\nu}
|0^+_i}
\nonumber\\
&- \frac{1}{4}
\sum_{j_1j_2}
\sum_{j_1'j_2'}
\sqrt{1+(-1)^J\delta_{j_1j_2}}
\sqrt{1+(-1)^J\delta_{j_1'j_2'}}
\nonumber \\
&\quad
\times
O_{\alpha}(j_1j_2j_1'j_2';2)
\nonumber\\
&\quad
\times
B_\pi(j_1j_2)B_{\nu}(j_1'j_2')
\braket{0^+_f| 
d^\+_\pi \cdot \tilde d_{\nu}
|0^+_i} \; .
\end{align} 
The coefficients $A_\rho(j)$ and $B_\rho(j_1j_2)$ 
are computed by the method of 
Frank and Van Isacker \cite{frank1982}, 
which was also employed in Ref.~\cite{barea2009}. 
The exact forms of these coefficients 
are found in Appendix~\ref{sec:abjr}.
Note that in the expressions $s^\+_\pi \cdot \tilde s_{\nu}$ and 
$d^\+_\pi \cdot \tilde d_{\nu}$ in (\ref{eq:mzn-1}) 
bosons are treated as like particles. 
If neutron (proton) bosons are like holes, 
the neutron annihilation (proton creation) operators 
should be replaced with the creation (annihilation) 
operators. 
The matrix elements 
$\braket{0^+_f| s^\+_\pi \cdot \tilde s_{\nu} |0^+_i}$ and 
$\braket{0^+_f| d^\+_\pi \cdot \tilde d_{\nu} |0^+_i}$
are calculated using the $0^+$ wave functions 
for the parent and daughter nuclei 
that are eigenfunctions 
of the mapped IBM-2 Hamiltonian.

%-----------------------------------------------------------
%
%       Axial PESs
%
%-----------------------------------------------------------
\begin{figure}[ht!]
\begin{center}
\includegraphics[width=\linewidth]{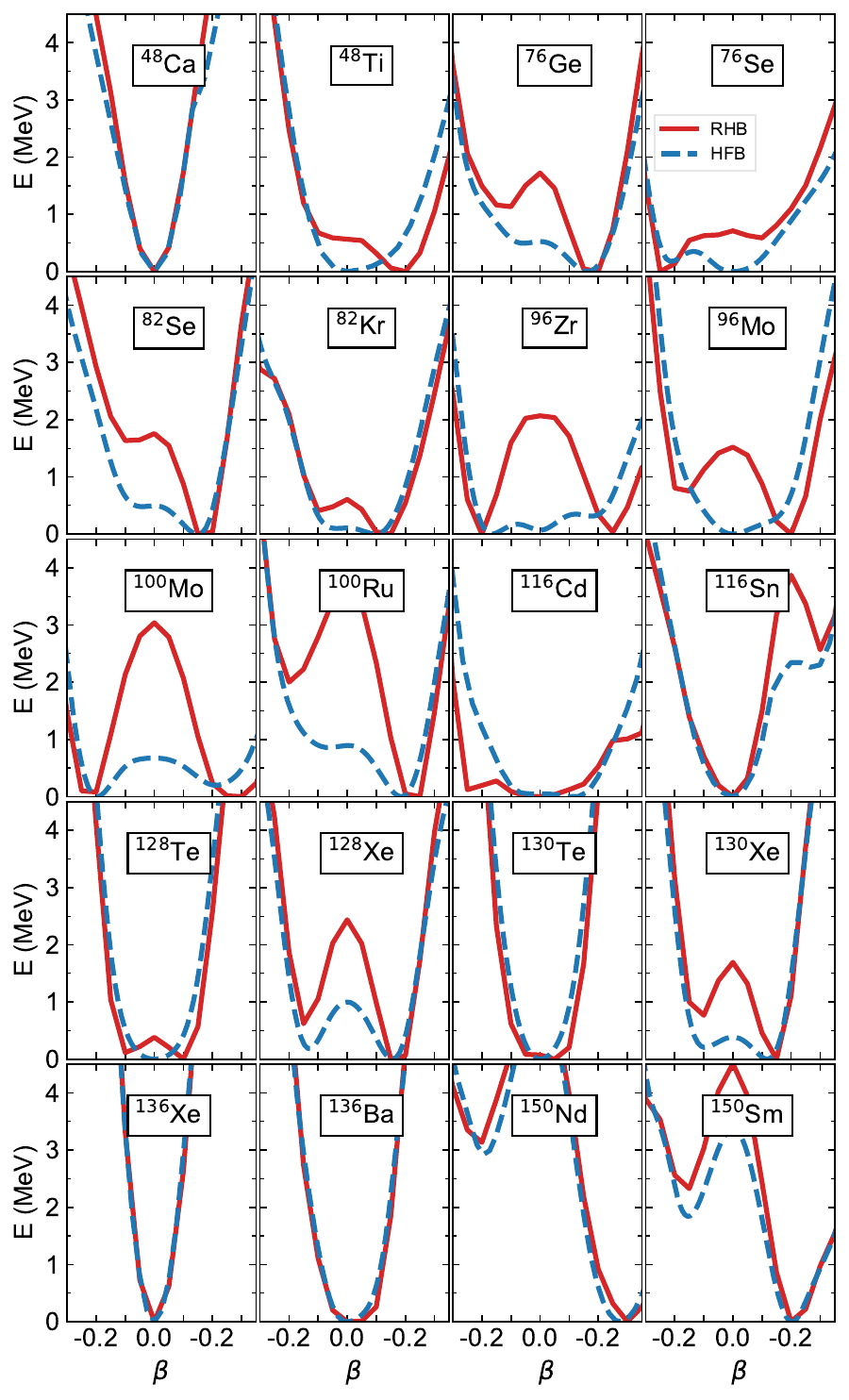}
\caption{
Potential energy curves along 
the $\beta$ deformation for the even-even nuclei 
involved in the $\znbb$-decay processes, 
computed by the constrained SCMF methods
within the RHB and within the HFB.
The total mean-field energies are 
plotted with respect to the global minimum. 
}
\label{fig:pes-axial}
\end{center}
\end{figure}

%-----------------------------------------------------------
%
%       IBM-2 yrast spectra
%
%-----------------------------------------------------------
\begin{figure}[ht]
\begin{center}
\includegraphics[width=\linewidth]{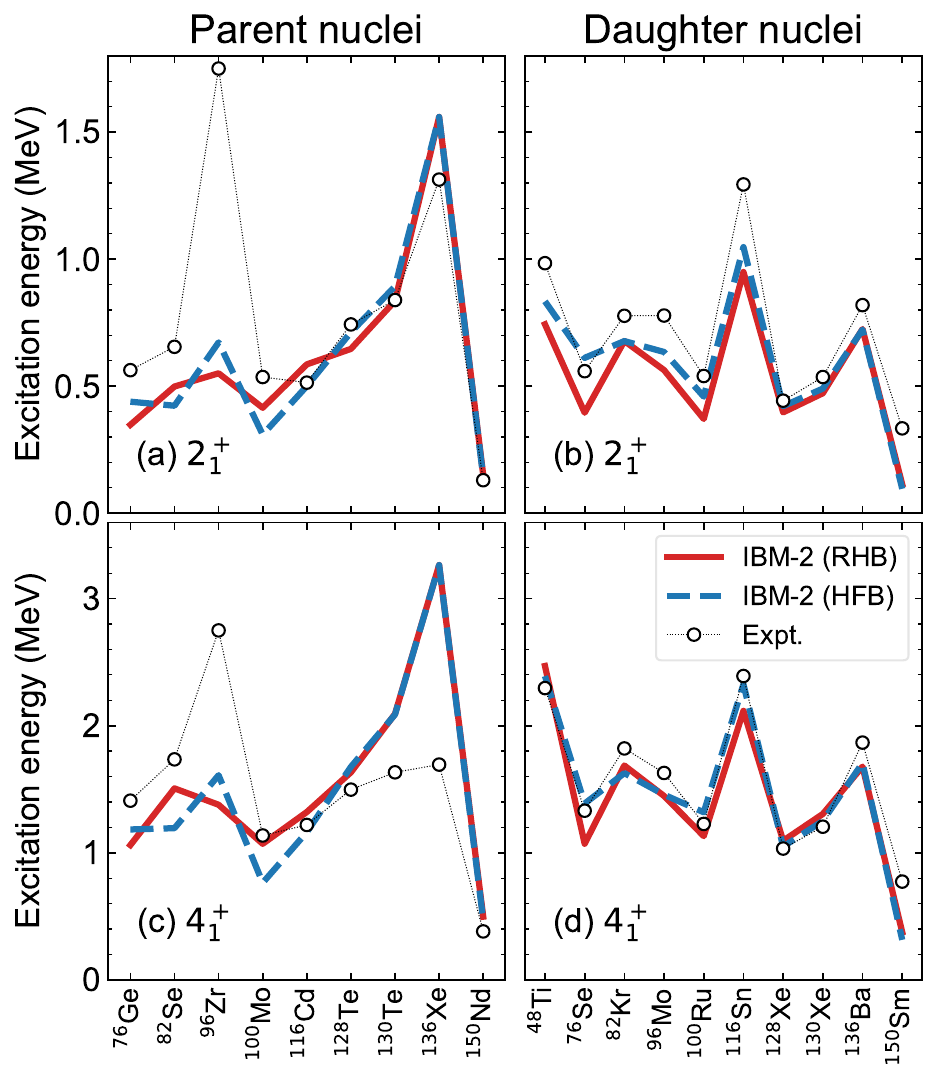}
\caption{
Excitation energies of the $2^+_1$
and $4^+_1$ states in the (a), (c) parent
and (b), (d) daughter
even-even nuclei of the $\znbb$ decays, 
obtained from the mapped IBM-2 calculations 
that are based on the RHB and HFB SCMF models.
Experimental data are taken from the
Brookhaven National Nuclear Data
Center \cite{data}. 
}
\label{fig:yrast}
\end{center}
\end{figure}

%-----------------------------------------------------------
%
%       IBM-2 non-yrast spectra
%
%-----------------------------------------------------------
\begin{figure}[ht]
\begin{center}
\includegraphics[width=\linewidth]{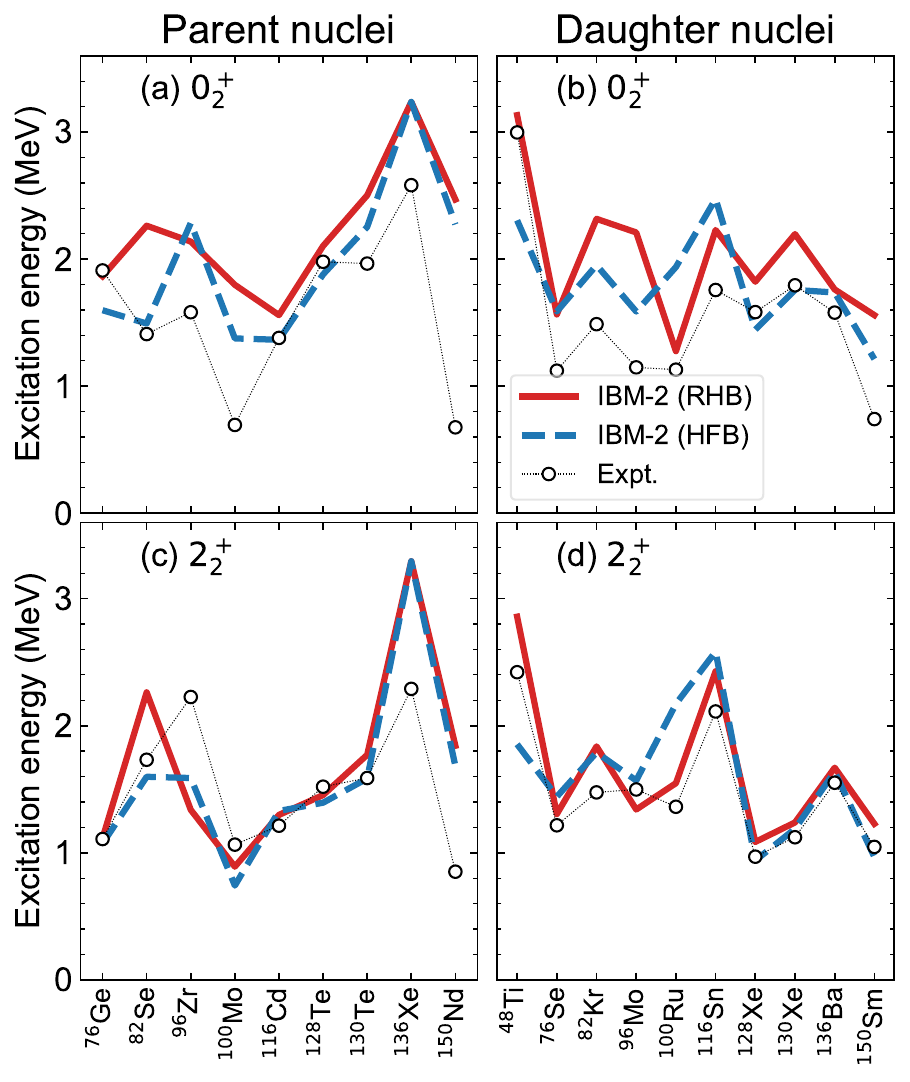}
\caption{
Same as the caption to Fig.~\ref{fig:yrast}, but for 
the $0^+_2$ and $2^+_2$ states.
}
\label{fig:nonyrast}
\end{center}
\end{figure}

%-----------------------------------------------------------
%
%       B(E2) for initial nuclei
%
%-----------------------------------------------------------
\begin{figure}[ht]
\begin{center}
\includegraphics[width=\linewidth]{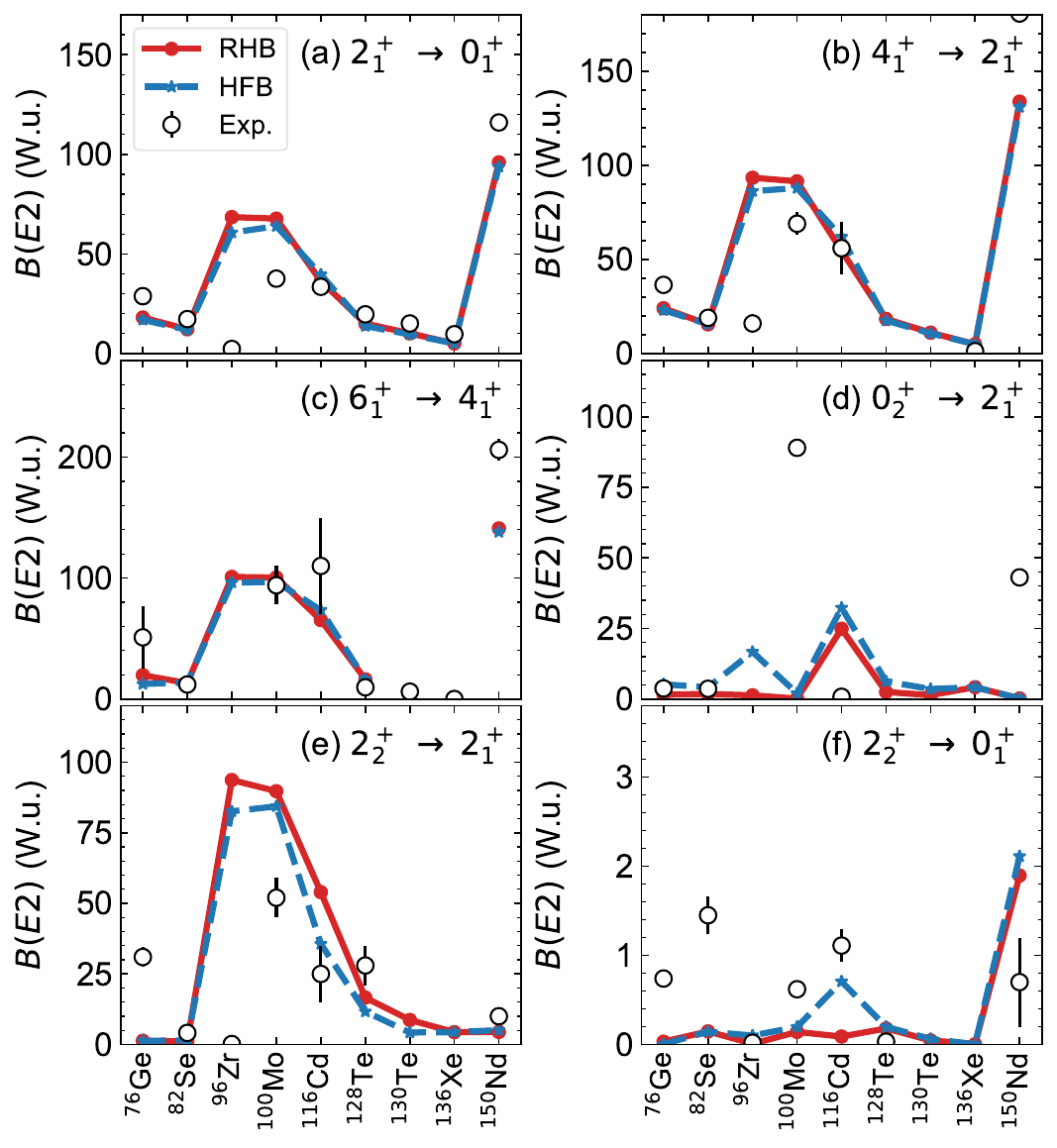}
\caption{
Calculated and experimental \cite{data} $B(E2)$
values for low-lying states in parent nuclei.
}
\label{fig:be2_ee1}
\end{center}
\end{figure}

%-----------------------------------------------------------
%
%       B(E2) for final nuclei
%
%-----------------------------------------------------------
\begin{figure}[ht]
\begin{center}
\includegraphics[width=\linewidth]{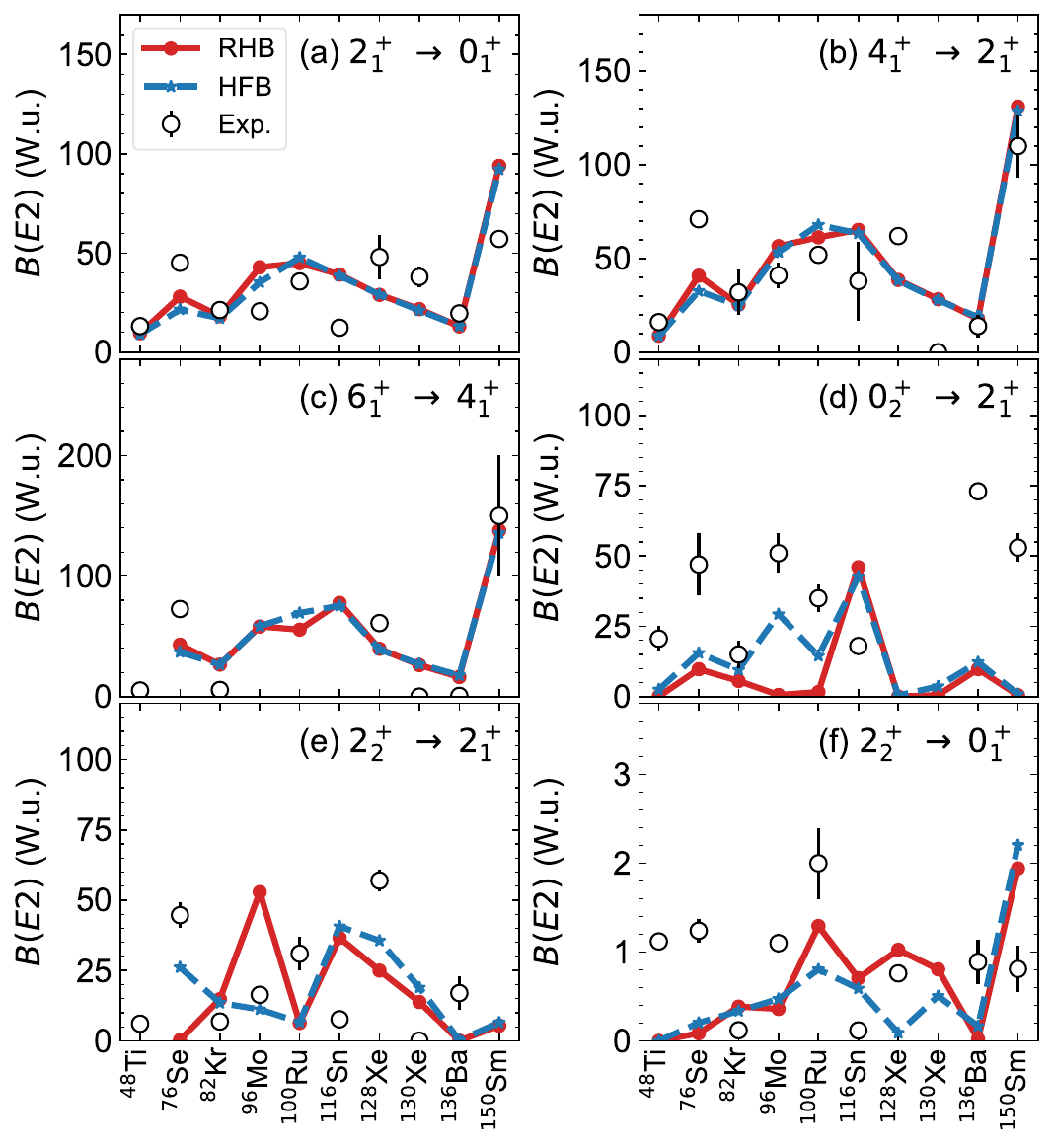}
\caption{
Same as the caption of Fig.~\ref{fig:be2_ee1},
but for daughter nuclei.
}
\label{fig:be2_ee2}
\end{center}
\end{figure}

%-----------------------------------------------------------
%
%       q and m moments
%
%-----------------------------------------------------------
\begin{figure}[ht]
\begin{center}
\includegraphics[width=\linewidth]{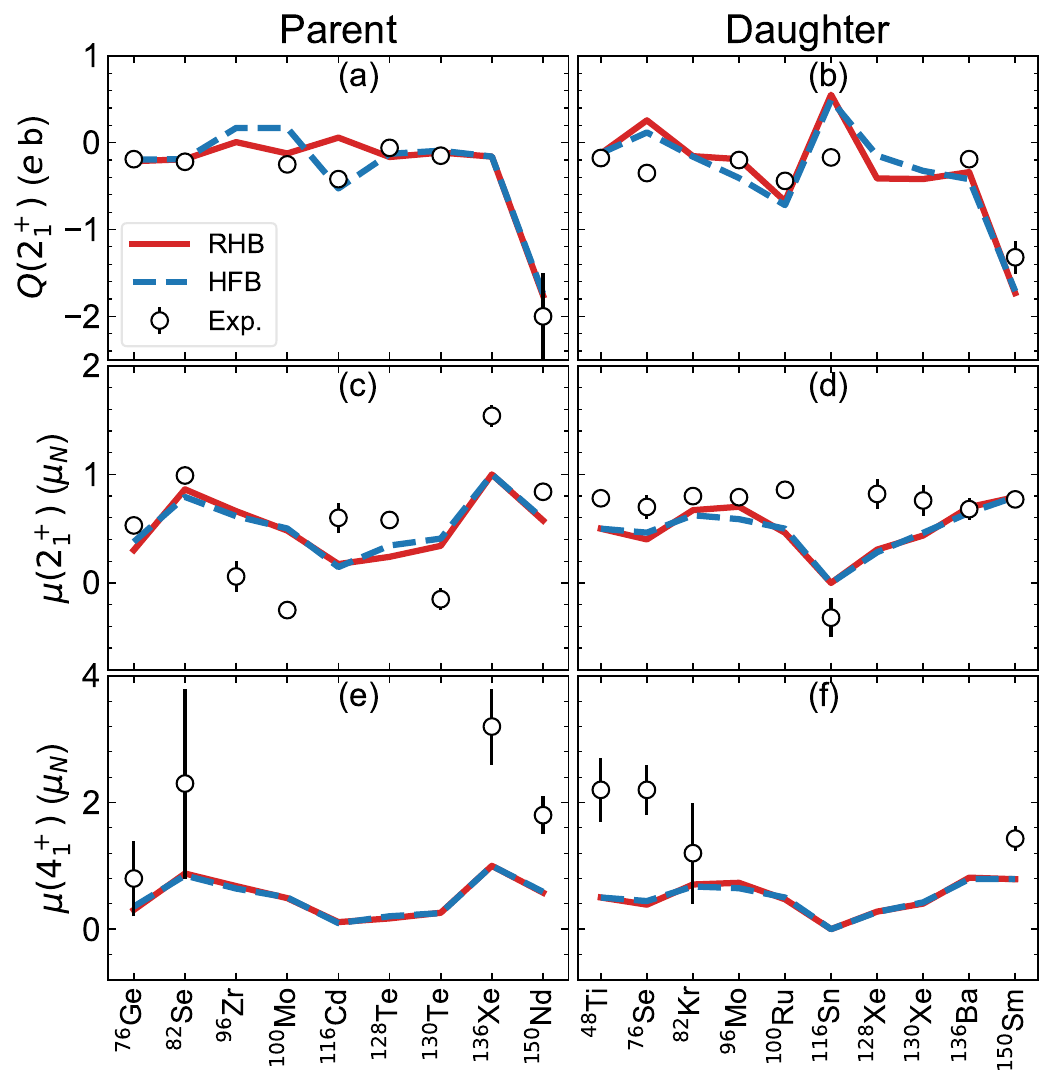}
\caption{
Electric quadrupole moments $Q(I)$ in $e$b units
for the $2^+_1$ state, and magnetic dipole moments $\mu(I)$ in
$\mu_N$ (nuclear magneton) units for
the $2^+_1$ and $4^+_1$ states
of the considered even-even nuclei.
The calculations are made within the RHB and HFB
frameworks, and the experimental values
are adopted from Ref.~\cite{data}.
}
\label{fig:mom}
\end{center}
\end{figure}

\section{Low-energy nuclear structures\label{sec:str}}

\subsection{Potential energy curves}

Figure~\ref{fig:pes-axial} shows potential energy curves 
as functions of the axial deformation $\beta$
computed for the even-even nuclei within the 
RHB and HFB methods.
The triaxial quadrupole $(\beta,\gamma)$ SCMF PESs
computed with the RHB method for the studied
even-even nuclei are found
in Refs.~\cite{nomura2022bb,nomura2024bb}. 
The HFB $(\beta,\gamma)$ SCMF PESs 
for most of the even-even nuclei are
taken from the previous studies:
Ref.~\cite{nomura2017ge} for $^{76}$Ge, $^{76}$Se, and $^{82}$Se;
Ref.~\cite{nomura2017kr} for $^{82}$Kr;
Ref.~\cite{nomura2016zr} for $^{96}$Zr, $^{96}$Mo, $^{100}$Mo, and $^{100}$Ru;
Ref.~\cite{nomura2017odd-3} for $^{128}$Xe, $^{130}$Xe and $^{136}$Ba;
Ref.~\cite{nomura2017odd-2} for $^{150}$Sm;
and Ref.~\cite{nomura2021oct-ba} for $^{150}$Nd.
The intrinsic and spectroscopic properties of these even-even 
nuclei related to the onset of deformations
and shape phase transitions have been discussed
in detail in the references mentioned above.
The HFB PESs
for the $^{48}$Ca, $^{48}$Ti, $^{116}$Cd, $^{116}$Sn, 
$^{128}$Te, $^{130}$Te and $^{136}$Xe nuclei,
shown in Fig.~\ref{fig:pes-axial},
are here computed by using the code HFBTHO \cite{hfbtho400}, 
which assumes the axial symmetry.

\subsection{Low-energy spectra}

Figure~\ref{fig:yrast} gives 
the excitation energies of the yrast 
states $2^+_1$ and $4^+_1$ of the 
even-even parent and daughter nuclei 
computed by the mapped IBM-2 based on the
RHB and HFB models.
The calculated excitation energies are 
consistent with the experimental data, 
except for $^{96}$Zr and $^{136}$Xe, for which 
the calculations considerably 
underestimate the experimental levels. 
The nucleus $^{96}$Zr corresponds to the 
neutron $N=56$ and proton $Z=40$ 
subshell closures, and 
its ground state has indeed been suggested  
to be spherical in nature experimentally \cite{kremer2016}. 
Both the relativistic and nonrelativistic 
SCMF calculations for this nucleus 
rather suggest the PES 
that exhibits a large prolate or 
oblate deformation (see Fig.~\ref{fig:pes-axial}) 
and, consequently, the mapped 
IBM-2 Hamiltonian produces 
unexpectedly low-lying $2^+_1$ and $4^+_1$ 
energy levels. 
Overestimates of the $4^+_1$ energy 
for $^{136}$Xe are 
due to the fact this nucleus corresponds 
to the neutron magic number $N=82$, in which case 
the IBM in general is not quite reliable 
because it is built only on the valence space.

One can also observe that
the IBM-2 spectra obtained from the
RHB calculations are quantitatively different from
those from the HFB calculations.
As shown in Fig.~\ref{fig:pes-axial} 
the RHB PESs generally exhibit 
a more pronounced deformed minimum or 
steeper potential valley than the HFB PESs.
The IBM-2 mapping from the RHB PES is, 
therefore, supposed to produce a more 
rotational-like energy spectrum characterized 
by the compressed energy levels than 
in the case of the HFB.

Calculation of the excited $0^+$ states is 
important, since the $\db$ decay from the 
ground-state $0^+_1$ to excited $0^+_2$ states 
is also possible. 
There is indeed experimental 
evidence for the $0^+_1$ $\to$ $0^+_2$ 
$\tnbb$ decays in $^{100}$Mo \cite{augier2023-100Mo} 
and $^{150}$Nd (see Ref.~\cite{barabash2020}
and references therein). 
The mapped IBM-2, particularly with the RHB input,
systematically overestimates the $0^+_2$ 
energy levels for parent and daughter nuclei. 
The overestimate of the $0^+_2$ levels 
appears to be a general feature 
of the mapped IBM-2 descriptions, 
which has occurred in different mass regions, 
and can be mainly attributed to the fact that 
the quadrupole-quadrupole boson interaction 
strength $\kappa$ derived from the mapping 
takes a very large negative value,
which is larger than those typically used in the
phenomenological IBM-2 calculations
by a factor of $\approx$ 2-4.
With the large negative $\kappa$ values, 
the $d$-boson contributions to low-lying states 
are appreciably large
and lead to a rotational-like 
level structure, in which the energy levels in 
the ground-state band are compressed
and the $0^+_2$ energy level 
appears at a relatively high energy. 
The value of the derived $\kappa$ parameter 
also reflects the features of the SCMF PES. 
In particular, the potential valley 
of the PES computed by the SCMF method with 
many of the relativistic and nonrelativistic 
EDFs appears to be often quite steep 
in both $\beta$ and $\gamma$ deformations. 
To reproduce such a feature, 
the mapping procedure requires to choose 
the large $|\kappa|$ values. 
The calculated energy levels, particularly 
those of the non-yrast states, 
depend on the feature of the corresponding PES. 
The RHB and HFB PESs indeed exhibit 
different topology in many cases 
(see Fig.~\ref{fig:pes-axial}). 
The HFB-mapped IBM-2 appears to 
give a lower $0^+_2$ excitation energy than 
the RHB-mapped IBM-2
(see Fig.~\ref{fig:nonyrast}).

In addition, 
in a few transitional nuclei, e.g., $^{100}$Mo, 
the observed $0^+_2$ energy levels 
are considerably low 
[see Figs.~\ref{fig:nonyrast}(a) and 
\ref{fig:nonyrast}(b)]. 
These extremely low-lying $0^+_2$ states 
may be attributed to the intruder particle-hole 
excitations, and cannot be reproduced by 
the standard IBM-2. 
The effects of the intruder states could be 
accounted for in the IBM-2 by extending it 
to include configuration mixing 
\cite{duval1981,nomura2012sc,nomura2016zr}. 

The level structure of the $\gamma$-soft systems 
is characterized by the low-lying $2^+_2$ state close 
in energy to the ground-state yrast band, 
and this state is often interpreted as the bandhead 
of the $\gamma$-vibrational band. 
In most cases, the mapped IBM-2 results shown in 
Figs.~\ref{fig:nonyrast}(c) and \ref{fig:nonyrast}(d) 
are consistent with the experimental $2^+_2$ levels, 
particularly for the daughter nuclei.
Certain deviations from the data are found for
$^{96}$Zr and $^{136}$Xe, because the IBM-2 does not
properly account for the (sub-)shell 
closure effects in these nuclei.

\subsection{Electromagnetic properties}

Electromagnetic transition properties are a stringent
test of the mapped IBM-2 wave functions.
The electric quadrupole $E2$ and magnetic dipole
$M1$ operators are defined as
\begin{align}
& \hat T^{E2} = e_{\rm B}^{\nu}\hat Q_{\nu}+
e_{\rm B}^{\pi}\hat Q_{\pi} \\
& \hat T^{M1} = \sqrt{\frac{3}{4\pi}}
\left(
g_{\rm B}^{\nu}\hat L_{\nu}+
g_{\rm B}^{\pi}\hat L_{\pi}
\right) \; ,
\end{align}
where $e_{\rm B}^{\rho}$, $\hat Q_{\rho}$,
$g_{\rm B}^{\rho}$, and $\hat L_{\rho}$ are
boson effective charge, quadrupole operator
same as that used in the Hamiltonian \eqref{eq:hb},
boson gyromagnetic ($g$) factor,
and the angular momentum operator
$\hat L_{\rho} = \sqrt{10}[d_{\rho}^{\+}\times\tilde d_{\rho}]^{(1)}$.
The neutron $e_{\rm B}^{\nu}$ and proton
$e_{\rm B}^{\pi}$ $E2$ effective charges are
here assumed to be equal,
$e_{\rm B}^{\nu}=e_{\rm B}^{\pi}\equiv e_{\rm B}$
and the fixed values are adopted for different mass
regions, that are similar to those
used in the earlier mapped IBM-2 calculations or
earlier microscopic IBM-2 studies:
$e_{\rm B}=0.06$ $e$b \cite{nomura2022beta-ge} for $A=48-82$, 
0.1 $e$b \cite{nomura2020zr} for $A=100-136$,
and 0.13 $e$b \cite{scholten1986} for $A=150$ regions.
For the bosonic $g$-factors, standard values
$g_{\rm B}^{\nu}=0$ $\mu_{N}$ and
$g_{\rm B}^{\pi}=1.0$ $\mu_{N}$
are adopted.

Figures~\ref{fig:be2_ee1} and \ref{fig:be2_ee2}
show the calculated $B(E2)$ values in
Weisskopf units (W.u.) for those transitions
between some low-lying states.
Note that there is no $6^+$ state
for $^{48}$Ti, $^{130}$Te,
and $^{136}$Xe due to limitations of the
boson configuration space, and the corresponding
$B(E2;6^+_1 \to 4^+_1)$ values are
not shown in the figures.
The calculated $B(E2)$ rates for the transitions
between the yrast states
$B(E2;2^+_1\to 0^+_1)$,
$B(E2;4^+_1\to 2^+_1)$, and
$B(E2;6^+_1 \to 4^+_1)$ are, in most cases,
of the same order of magnitudes as
the experimental values \cite{data}.
It should be noticed that the mapped IBM-2
results significantly overestimate
the observed $B(E2;2^+_1\to 0^+_1)$ and
$B(E2;4^+_1\to 2^+_1)$ values for $^{96}$Zr.
The experimental
$B(E2;2^+_1\to 0^+_1)$ rate is,
however, also negligibly
small, and does not represent
a strong collectivity.
To reproduce the $B(E2)$ systematic
in $^{96}$Zr some additional correlations
may need to be introduced in
the IBM-2 mapping procedure,
such as the configuration mixing.
For the nuclei $^{82}$Kr, $^{130}$Xe,
$^{136}$Xe and $^{136}$Ba, the experimental
$B(E2;6^+_1 \to 4^+_1)$ values are smaller
than the $B(E2;4^+_1\to 2^+_1)$ values and,
in some cases, the $B(E2;4^+_1\to 2^+_1)$ values
are suggested to be even smaller than the
$B(E2;2^+_1\to 0^+_1)$ values.
The present calculations do not reproduce
this systematic, but these discrepancies are not surprising,
since the yrast states $4^+_1$ and $6^+_1$ for
these nuclei are supposed to have properties
that cannot be described by the standard IBM-2,
which gives increasing $B(E2)$ rates within the
ground-state band as functions of spin.

The mapped IBM-2 calculations overall result in
small $B(E2;0^+_2\to 2^+_1)$ values, underestimating
experimental values for many of the studied nuclei.
The small $0^+_2\to 2^+_1$ transition strengths
indicate a weak coupling of the $0^+_2$
state to the $2^+_1$ state.
Within the present framework, this occurs
because the underlying SCMF PESs have
a pronounced energy minimum, and
the resulting IBM-2 spectrum has characteristics
that resemble those in the rotational limit, in which
the $0^+_2\to 2^+_1$ transition is weak.
The predicted $B(E2;2^+_2\to 2^+_1)$ rates,
resulting either from the RHB or HFB inputs,
are more or less of the same order of
magnitudes as the experimental values.
A few exceptions are found in $^{76}$Ge,
$^{76}$Se (with the RHB input), and $^{96}$Zr.
The observed $B(E2;2^+_2\to 2^+_1)$ values
for $^{76}$Ge and $^{76}$Se , in particular,
are so large as to be of the same order
of magnitude as the $B(E2;2^+_1\to 0^+_1)$ values,
indicating pronounced $\gamma$ softness.
The mapped IBM-2 significantly underestimates the
experimental $B(E2;2^+_2\to 2^+_1)$ values
for these nuclei.
The calculated values for the
$B(E2;2^+_2\to 0^+_1)$ transition
strengths are also shown in Figs.~\ref{fig:be2_ee1}
and \ref{fig:be2_ee2}.
The calculated $B(E2;2^+_2\to 0^+_1)$
values are generally consistent with
the experimental values.

Note that the calculated values
for the $B(E2;2^+_1\to 0^+_1)$,
$B(E2;4^+_1\to 2^+_1)$, and
$B(E2;6^+_1 \to 4^+_1)$ transition rates with the
RHB input do not significantly differ from those
with the HFB input.
Some differences between the two sets of
the calculations appear in the $B(E2)$ values
for the transitions that
involve non-yrast states in a few nuclei.

Figure~\ref{fig:mom} displays the predicted
(spectroscopic) quadrupole moment $Q(2^+_1)$ in $e$b
for the $2^+_1$ state, and magnetic dipole moments
$\mu(2^+_1)$ and $\mu(4^+_1)$ in nuclear magneton ($\mu_N$).
The mapped IBM-2 calculations with both
the RHB and HFB inputs produce the
$Q(2^+_1)$ values, including sign, that
are consistent with the experimental values \cite{data}.
Positive $Q(2^+_1)$ values for $^{76}$Se in the
IBM-2, which contradict the experimental data,
reflect the oblate equilibrium minimum
in the potential energy curves.
The calculated $\mu(2^+_1)$ moments are
also overall consistent with the observed values \cite{data},
both in magnitude and sign.
The $\mu(4^+_1)$ moments are here predicted to
be systematically lower than data but have
the correct sign.

\section{$\znbb$ decay\label{sec:db}}

%-----------------------------------------------------------
%
%     NME table
%
%-----------------------------------------------------------
\begin{table*}
\caption{\label{tab:nme}
Predicted $\mgt$, $\mfe$, 
and $\mte$ matrix elements, and total NME $\mzn$ 
for the $0^+_1 \to 0^+_1$ $\znbb$ decays within 
the mapped IBM-2 based on 
the RHB and HFB SCMF calculations. 
}
 \begin{center}
 \begin{ruledtabular}
  \begin{tabular}{lcccccccc}
\multirow{2}{*}{Decay} & \multicolumn{4}{c}{RHB} &
\multicolumn{4}{c}{HFB} \\
\cline{2-5}\cline{6-9}
& $\mgt$ & $\mfe$ & $\mte$ & $\mzn$ & $\mgt$ & $\mfe$ & $\mte$ & $\mzn$ \\
\hline
$^{48}$Ca $\to$ $^{48}$Ti & $0.713$ & $-0.035$ & $-0.075$ & $0.659$ & $1.760$ & $-0.062$ & $-0.081$ & $1.717$ \\
$^{76}$Ge $\to$ $^{76}$Se & $2.801$ & $-0.107$ & $-0.061$ & $2.806$ & $4.066$ & $-0.163$ & $-0.123$ & $4.045$ \\
$^{82}$Se $\to$ $^{82}$Kr & $2.136$ & $-0.089$ & $-0.062$ & $2.130$ & $2.687$ & $-0.113$ & $-0.094$ & $2.664$ \\
$^{96}$Zr $\to$ $^{96}$Mo & $3.258$ & $-0.419$ & $0.143$ & $3.662$ & $3.636$ & $-0.558$ & $0.141$ & $4.123$ \\
$^{100}$Mo $\to$ $^{100}$Ru & $2.732$ & $-0.263$ & $0.143$ & $3.038$ & $3.490$ & $-0.439$ & $0.157$ & $3.920$ \\
$^{116}$Cd $\to$ $^{116}$Sn & $3.379$ & $-0.564$ & $0.130$ & $3.859$ & $3.479$ & $-0.549$ & $0.138$ & $3.958$ \\
$^{128}$Te $\to$ $^{128}$Xe & $1.671$ & $-0.087$ & $-0.024$ & $1.702$ & $2.986$ & $-0.161$ & $-0.060$ & $3.026$ \\
$^{130}$Te $\to$ $^{130}$Xe & $1.474$ & $-0.076$ & $-0.019$ & $1.502$ & $3.016$ & $-0.163$ & $-0.063$ & $3.055$ \\
$^{136}$Xe $\to$ $^{136}$Ba & $2.082$ & $-0.113$ & $-0.047$ & $2.106$ & $2.545$ & $-0.139$ & $-0.061$ & $2.570$ \\
$^{150}$Nd $\to$ $^{150}$Sm & $4.262$ & $-0.844$ & $0.142$ & $4.929$ & $4.078$ & $-0.923$ & $0.111$ & $4.762$ \\
  \end{tabular}
 \end{ruledtabular}
 \end{center}
\end{table*}

%-----------------------------------------------------------
%
%     NME table 0^+_2
%
%-----------------------------------------------------------
\begin{table*}
\caption{\label{tab:nme-ex}
Same as the caption to Table~\ref{tab:nme}, but
for the $0^+_1 \to 0^+_2$ $\znbb$ decays.}
 \begin{center}
 \begin{ruledtabular}
  \begin{tabular}{lcccccccc}
\multirow{2}{*}{Decay} & \multicolumn{4}{c}{RHB} &
\multicolumn{4}{c}{HFB} \\
\cline{2-5}\cline{6-9}
& $\mgt$ & $\mfe$ & $\mte$ & $\mzn$ & $\mgt$ & $\mfe$ & $\mte$ & $\mzn$ \\
\hline
$^{48}$Ca $\to$ $^{48}$Ti & $2.743$ & $-0.086$ & $-0.051$ & $2.745$ & $2.430$ & $-0.067$ & $-0.012$ & $2.460$ \\
$^{76}$Ge $\to$ $^{76}$Se & $1.300$ & $-0.054$ & $-0.053$ & $1.280$ & $0.314$ & $-0.016$ & $-0.037$ & $0.286$ \\
$^{82}$Se $\to$ $^{82}$Kr & $0.584$ & $-0.021$ & $-0.040$ & $0.557$ & $0.742$ & $-0.028$ & $-0.047$ & $0.712$ \\
$^{96}$Zr $\to$ $^{96}$Mo & $0.250$ & $-0.033$ & $0.011$ & $0.281$ & $0.347$ & $-0.052$ & $0.014$ & $0.393$ \\
$^{100}$Mo $\to$ $^{100}$Ru & $0.093$ & $-0.011$ & $0.004$ & $0.104$ & $0.978$ & $-0.125$ & $0.043$ & $1.099$ \\
$^{116}$Cd $\to$ $^{116}$Sn & $0.607$ & $-0.104$ & $0.017$ & $0.689$ & $0.488$ & $-0.069$ & $0.015$ & $0.545$ \\
$^{128}$Te $\to$ $^{128}$Xe & $0.139$ & $-0.007$ & $-0.003$ & $0.141$ & $0.934$ & $-0.050$ & $-0.022$ & $0.943$ \\
$^{130}$Te $\to$ $^{130}$Xe & $1.018$ & $-0.055$ & $-0.025$ & $1.026$ & $2.649$ & $-0.142$ & $-0.067$ & $2.671$ \\
$^{136}$Xe $\to$ $^{136}$Ba & $1.477$ & $-0.078$ & $-0.043$ & $1.483$ & $2.002$ & $-0.105$ & $-0.056$ & $2.010$ \\
$^{150}$Nd $\to$ $^{150}$Sm & $0.748$ & $-0.089$ & $0.026$ & $0.829$ & $0.441$ & $-0.073$ & $0.007$ & $0.494$ \\
  \end{tabular}
 \end{ruledtabular}
 \end{center}
\end{table*}

%-----------------------------------------------------------
%
%     NMEs in comparison with other predictions
%
%-----------------------------------------------------------
\begin{figure*}[ht!]
\begin{center}
\includegraphics[width=\linewidth]{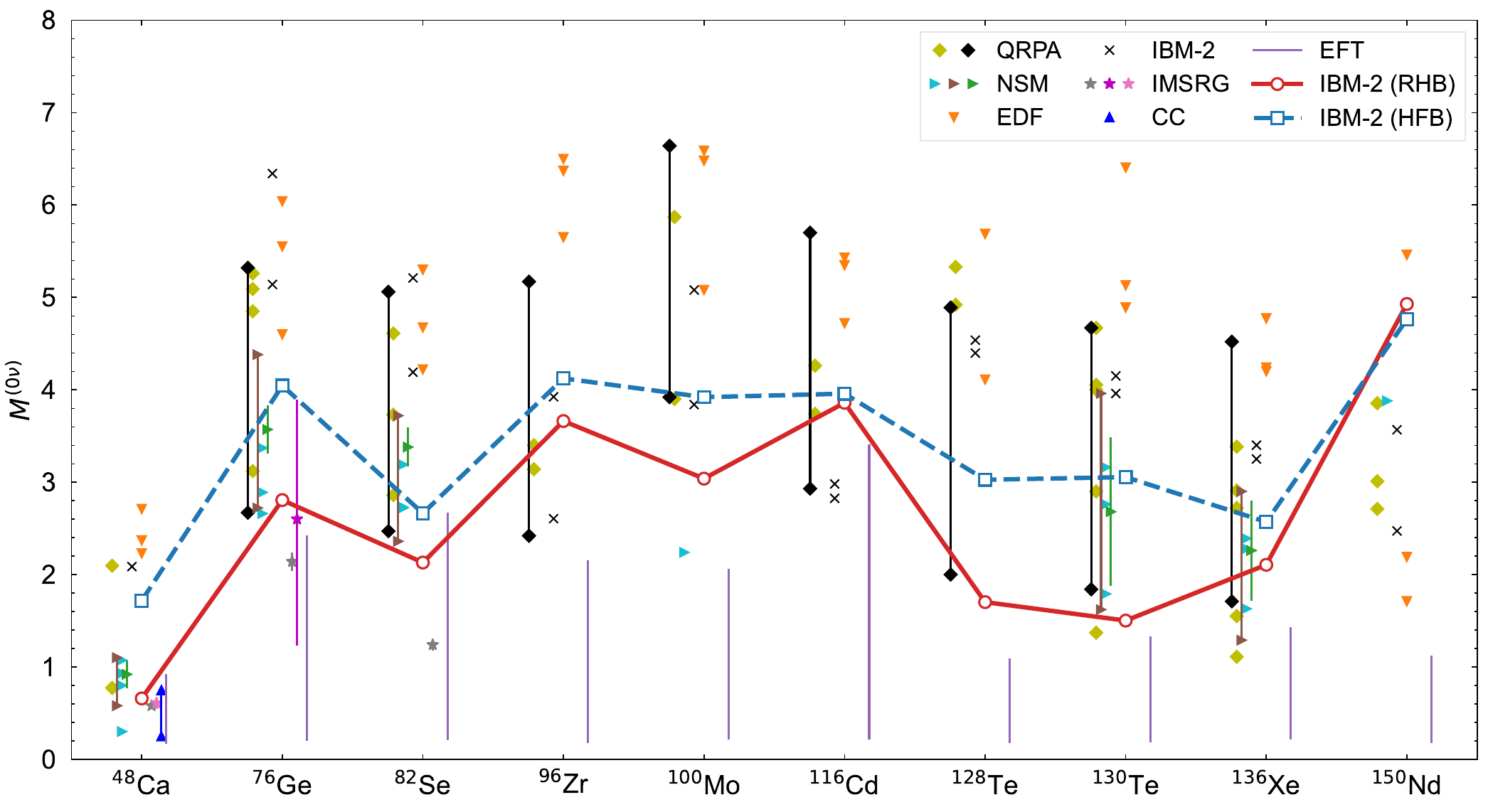}
\caption{
NMEs $\mzn$ for
the $0^+_1\,\to\,0^+_1$ $\znbb$ decays
of the candidate nuclei of interest,
obtained from
the mapped IBM-2
calculations based on the RHB
and HFB SCMF models.
Those NMEs from other many-body
methods are also shown:
QRPA (yellow \cite{mustonen2013,simkovic2018,fang2018,hyvarinen2015,terasaki2020}, and black \cite{jokiniemi2023} diamonds),
NSM (cyan \cite{horoi2016,iwata2016,menendez2018,corragio2020,tsunoda2023},
brown \cite{jokiniemi2023}, and
green \cite{castillo2025} right triangles),
EDF-GCM \cite{trodriguez2010,vaquero2013,song2017}, 
IBM-2 \cite{barea2015,deppisch2020}, 
IMSRG (gray \cite{yao2020}, magenta \cite{belley2021},
and pink \cite{belley2024} stars), 
CC \cite{novario2021}, and EFT \cite{brase2022}.}
\label{fig:nme}
\end{center}
\end{figure*}

%-----------------------------------------------------------
%
%     <ss>, <dd>
%
%-----------------------------------------------------------
\begin{figure}[ht]
\begin{center}
\includegraphics[width=\linewidth]{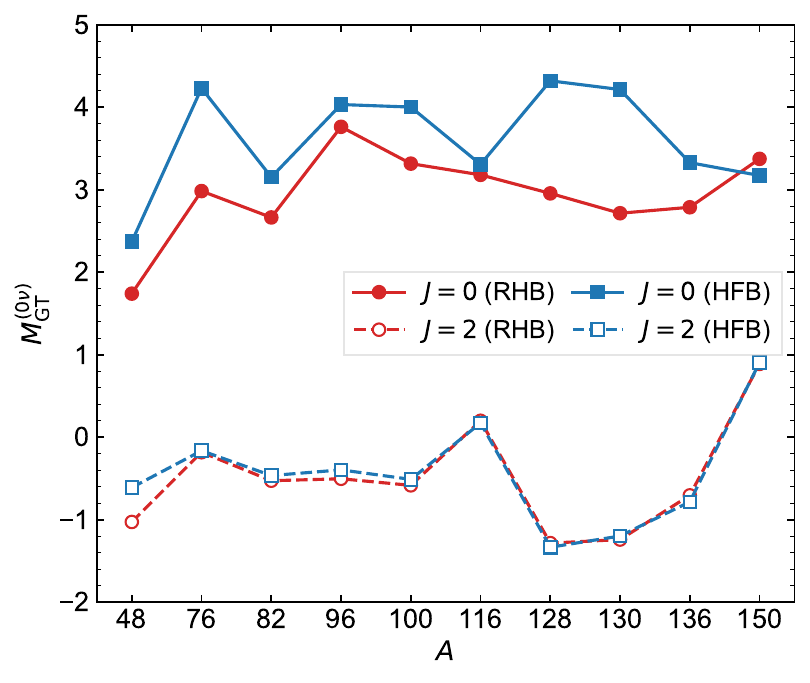}
\caption{Decomposition of the GT matrix elements 
into the monopole ($J=0$) and 
quadrupole ($J=2$) components.}
\label{fig:ssdd}
\end{center}
\end{figure}

Predicted $\mgt$, $\mfe$ and $\mte$
matrix elements, 
and final NMEs $\mzn$ for the $0^+_1 \to 0^+_1$ decays  
for the studied $\znbb$ emitters are 
shown in Table~\ref{tab:nme}. 
Two sets of results shown in the table correspond 
to the calculations employing the RHB and HFB methods
for the self-consistent calculations. 
The dominant contributions to the total NMEs $\mzn$ 
are from the GT transitions, while the Fermi and tensor 
parts appear to play less significant roles. 
In addition, the NMEs from the HFB are generally 
larger than those from the RHB.

The NMEs for the
$0^+_1$ $\to$ $0^+_2$ $\znbb$ decays 
are also computed with the mapped IBM-2, 
and appear to depend on the choice of the
microscopic input (see Table~\ref{tab:nme-ex}).
The $0^+_1$ $\to$ $0^+_2$ decay rates are particularly 
sensitive to the description of the 
$0^+_2$ states of the final nuclei, since the 
predicted $0^+_2$ energy levels for the daughter 
nuclei depend largely on whether the relativistic
or nonrelativistic SCMF is chosen
[see Fig.~\ref{fig:nonyrast}(b)]. 
In addition, the coexistence of more than one 
minimum observed in the PESs for several nuclei 
should have certain influences on the 
ground and excited $0^+$ states. 
For $^{96}$Zr, for instance, both the HFB
and RHB SCMF calculations suggest
two minima that are quite close in energy
\cite{nomura2016zr,nomura2022bb,nomura2024bb}. 
In such systems, substantial amounts of
shape mixing are supposed to be present in the IBM-2
lowest-lying $0^+$ states. 
A possible effect of the coexisting 
mean-field minima is discussed in Sec.~\ref{sec:co}.

Figure~\ref{fig:nme} displays 
the calculated $0^+_1$ $\to$ $0^+_1$ 
$\znbb$-decay NMEs, 
which are also shown in Table~\ref{tab:nme}, 
and those NMEs values computed by the 
QRPA \cite{mustonen2013,hyvarinen2015,simkovic2018,fang2018,terasaki2020,jokiniemi2023}, 
NSM \cite{horoi2016,iwata2016,menendez2018,corragio2020,tsunoda2023,jokiniemi2023,castillo2025},
EDF-GCM \cite{trodriguez2010,vaquero2013,song2017}, 
IBM-2 \cite{barea2015,deppisch2020},
IMSRG \cite{yao2020,belley2021,belley2024},
CC \cite{novario2021}, and EFT \cite{brase2022}. 
The RHB-mapped IBM-2 NMEs for the
$^{48}$Ca decay are
small, $\mzn<1$, and are
close to the values
obtained from the NSM calculations. 
The HFB-mapped IBM-2 calculation gives a larger NME for the
$^{48}$Ca decay, being rather close to the 
earlier IBM-2 value \cite{barea2015}.
The two sets of the IBM-2 results differ significantly,
probably because the present HFB calculation
suggests a spherical minimum for the $^{48}$Ti,
whereas the RHB PES predicts
a deformed minimum at $\beta\approx 0.15$ 
(see Fig.~\ref{fig:pes-axial}).
For the $^{76}$Ge $\to$ $^{76}$Se decay, the 
RHB-mapped IBM-2 calculation gives the NME
that is more or less close
to the predictions from the IMSRG 
\cite{belley2021,belley2024}.
The HFB-mapped IBM-2, however, 
produces the much larger NME, being closer
to the QRPA values. 
The mapped IBM-2 NMEs for 
the $^{82}$Se $\to$ $^{82}$Kr decay are 
smaller than many of the calculated NMEs in the 
EDF, QRPA, NSM, and IBM-2, but are 
close to the IMSRG \cite{belley2021} 
and EFT \cite{brase2022} values. 
For both the $^{76}$Ge $\to$ $^{76}$Se and 
$^{82}$Se $\to$ $^{82}$Kr decays, the mapped IBM-2 NMEs 
are lower than those of the previous IBM-2
calculations \cite{barea2015,deppisch2020}
approximately by a factor of 2.

For the $^{96}$Zr $\to$ $^{96}$Mo, 
$^{100}$Mo $\to$ $^{100}$Ru, 
and $^{116}$Cd $\to$ $^{116}$Sn decays, 
the mapped IBM-2 yields the NMEs that are
more or less close to the 
IBM-2 values of Refs.~\cite{barea2015,deppisch2020}. 
The present values of the NMEs for the 
$^{128}$Te $\to$ $^{128}$Xe and 
$^{130}$Te $\to$ $^{130}$Xe decays 
are systematically smaller than those in 
the majority of the other model calculations. 
The RHB- and HFB-mapped IBM-2 NMEs also
differ for the above two decay processes.
For the $^{136}$Xe $\to$ $^{136}$Ba $\znbb$ decay, 
the two sets of the mapped IBM-2 calculations 
provide the NME values close to those from other approaches.
In contrast with all the other $\znbb$-decay processes,
the present NME values for the
$^{150}$Nd $\to$ $^{150}$Sm decay appear 
to be among the largest of the 
NME values found in the literature.

To give further insights into the nature of the 
calculated NMEs, 
the GT matrix element $\mgt$, a dominant 
factor in the total NME, is decomposed into 
monopole and quadrupole components, which correspond 
to the first and second terms in Eq.~(\ref{eq:mzn-1}), 
respectively. 
For the $0^+_1$ $\to$ $0^+_1$ $\znbb$-decay 
processes, the monopole contribution 
is dominant over the quadrupole one, 
as shown in Fig.~\ref{fig:ssdd}. 
In general, larger monopole contributions 
are obtained when the HFB framework
is adopted as a microscopic basis than 
in the case of the RHB one.
The quadrupole contributions of $\mgt$ obtained 
from the RHB and HFB do not differ, except for the
$^{48}$Ca decay.
The ratios of the quadrupole to monopole GT matrix elements 
calculated within the RHB are, therefore,
systematically larger than those with the HFB. 
For the $^{48}$Ca, $^{128}$Te and $^{130}$Te 
$\znbb$ decays, the quadrupole-to-monopole ratios 
amount to 59 \%, 43 \%, and 46 \%, respectively, 
in the case of the RHB input. 
In the calculations with the HFB,
these ratios are less than
30 \% for all the studied
$\znbb$-decay processes.
In the previous IBM-2 calculations,
there was a very large monopole
and a minor quadruple pair contributions
to the GT matrix element,
e.g., for the $^{76}$Ge decay
\cite{barea2009}, leading to a
much larger NME than the present value.
Also pair contributions
of higher multipoles than
quadrupole $J=2^+$ and monopole
$J=0^+$ to the GT matrix elements
could have impacts on the final NMEs.
These higher-order
contributions are included
in the NSM calculations, but have not
been in the IBM,
except perhaps for Ref.~\cite{vanisacker2017},
in which isoscalar pairs (bosons)
were incorporated in the
calculations for the $\znbb$
decays in the $^{48}$Ca region.
Extensions of the present IBM-2
mapping to include
these multipole pair effects
are an interesting open problem.

Table~\ref{tab:tau} gives 
the half-lives for the $0^+_1 \to 0^+_1$ $\znbb$ 
decays (\ref{eq:tau}), 
computed by using the NMEs shown in Table~\ref{tab:nme} and 
Fig.~\ref{fig:nme}. 
The phase-space factors $G_{0\nu}$ are adopted 
from Ref.~\cite{kotila2012}, and 
the average light neutrino mass 
of $\braket{m_{\nu}}=1$ eV is assumed. 
The upper limits of $\braket{m_{\nu}}$ 
estimated by using the current limits on the $\taubb$, 
adopted from the recent compilation 
of G\'omez-Cadenas {\it et al.} \cite{gomezcadenas2024}, 
are also shown in Table~\ref{tab:nme}, 
and it appears that the HFB-mapped IBM-2 
overall gives shorter $\taubb$, hence slightly 
more stringent limits on neutrino mass, than the RHB-mapped 
IBM-2 calculation.

%-----------------------------------------------------------
%
%     T_1/2 table
%
%-----------------------------------------------------------
\begin{table*}
\caption{\label{tab:tau}
Predicted half-lives $\taubb$ (in yr) 
for the $0^+_1 \to 0^+_1$ 
$\znbb$ decays within the mapped IBM-2 
assuming the average light neutrino mass 
of $\braket{m_{\nu}}=1$ eV. 
The results obtained 
with the microscopic inputs provided by the RHB
(column 2), and HFB (column 4) SCMF are shown. 
In columns 3 and 5 
shown are the upper limits of the 
neutrino mass estimated using  
the 90 \% C.L. limits 
on $\taubb$, adopted from the complication 
of Ref.~\cite{gomezcadenas2024} (column 6).
}
 \begin{center}
 \begin{ruledtabular}
  \begin{tabular}{lccccc}
\multirow{2}{*}{Decay} & 
\multicolumn{2}{c}{RHB} &
\multicolumn{2}{c}{HFB} & 
\multirow{2}{*}{$T_{1/2,{\rm expt}}^{(\znbb)}$ (yr)} \\
\cline{2-3}\cline{4-5}
& 
$\taubb$ (yr) & $\braket{m_{\nu}}$ (eV) &
$\taubb$ (yr) & $\braket{m_{\nu}}$ (eV) \\
\hline
$^{48}$Ca $\to$ $^{48}$Ti & $9.34\times10^{24}$ & $<12.690$ & $1.38\times10^{24}$ & $<4.860$     & $>5.8\times10^{22}$ \\
$^{76}$Ge $\to$ $^{76}$Se & $5.41\times10^{24}$ & $<0.173$ & $2.60\times10^{24}$ & $<0.120$     & $>1.8\times10^{26}$ \\
$^{82}$Se $\to$ $^{82}$Kr & $2.18\times10^{24}$ & $<0.789$ & $1.40\times10^{24}$ & $<0.630$     & $>4.6\times10^{24}$ \\
$^{96}$Zr $\to$ $^{96}$Mo & $3.65\times10^{23}$ & $<6.255$ & $2.88\times10^{23}$ & $<5.517$     & $>9.2\times10^{21}$ \\
$^{100}$Mo $\to$ $^{100}$Ru & $6.85\times10^{23}$ & $<0.673$ & $4.12\times10^{23}$ & $<0.516$     & $>1.8\times10^{24}$ \\
$^{116}$Cd $\to$ $^{116}$Sn & $4.05\times10^{23}$ & $<1.348$ & $3.85\times10^{23}$ & $<1.314$     & $>2.2\times10^{23}$ \\
$^{128}$Te $\to$ $^{128}$Xe & $5.92\times10^{25}$ & $<6.280$ & $1.87\times10^{25}$ & $<3.531$     & $>3.6\times10^{24}$ \\
$^{130}$Te $\to$ $^{130}$Xe & $3.14\times10^{24}$ & $<0.377$ & $7.59\times10^{23}$ & $<0.185$     & $>2.2\times10^{25}$ \\
$^{136}$Xe $\to$ $^{136}$Ba & $1.56\times10^{24}$ & $<0.375$ & $1.05\times10^{24}$ & $<0.307$     & $>2.3\times10^{26}$ \\
$^{150}$Nd $\to$ $^{150}$Sm & $6.58\times10^{22}$ & $<1.732$ & $7.04\times10^{22}$ & $<1.732$     & $>2.0\times10^{22}$ \\
  \end{tabular}
 \end{ruledtabular}
 \end{center}
\end{table*}

%-----------------------------------------------------------
%
%     A=76 NME parameter variation
%
%-----------------------------------------------------------
\begin{figure}[ht]
\begin{center}
\includegraphics[width=\linewidth]{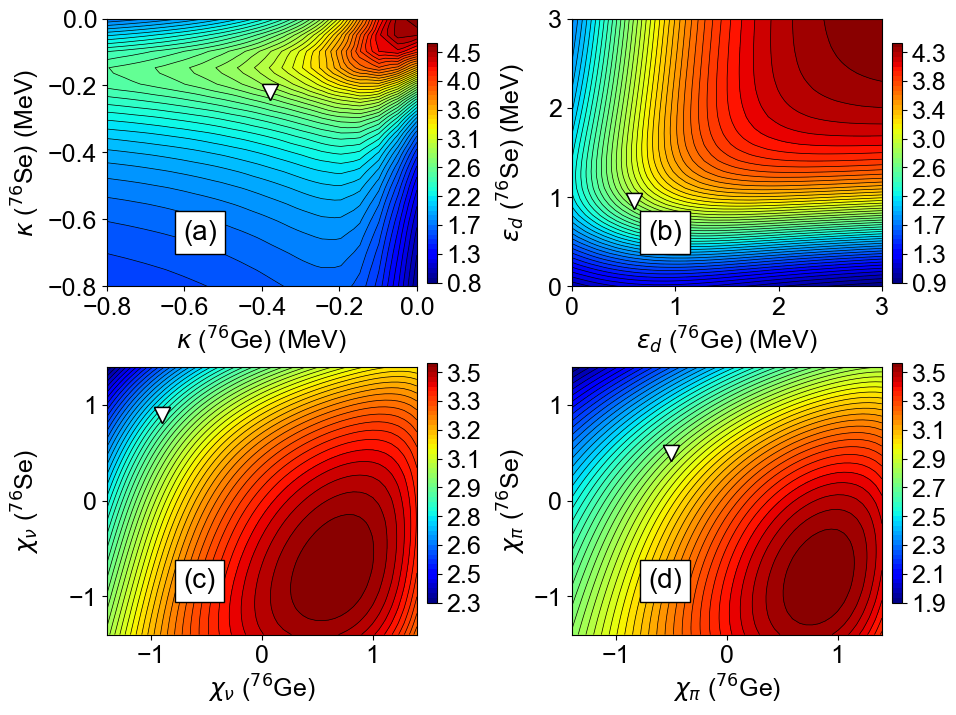}
\caption{ 
Calculated NMEs for the $\znbb$ decay
$^{76}$Ge$(0^+_1)$ $\to$ $^{76}$Se$(0^+_1)$ 
as functions of the parameters (a) $\kappa$, 
(b) $\epsilon_d$, (c) $\chi_{\nu}$, and (d) $\chi_{\pi}$
for the parent and daughter nuclei. 
The microscopic input 
to the IBM-2 mapping is based on
the RHB SCMF calculation.
The open triangle indicates the NME values 
calculated with the sets of the parameters  
obtained from the mapping.}
\label{fig:para-76gese}
\end{center}
\end{figure}

\section{Sensitivity analyses\label{sec:model}}

The predicted NME values appear to be 
sensitive to the parameters and assumptions 
considered in the calculations. 
In particular, 
it has been shown in preceding sections that 
the choice of the EDF considerably affects 
the energy spectra (Figs.~\ref{fig:yrast} 
and \ref{fig:nonyrast}) and $\znbb$-decay NMEs 
(Fig.~\ref{fig:nme}). 
The present section concerns discussions about
dependencies of the mapped IBM-2 NME 
results on the strength parameters and form
of the IBM-2 Hamiltonian, on the mapping procedure,
on the coexistence of mean-field minima
in the SCMF PESs that appears in some nuclei, 
and on the pair structure constants for 
the $\znbb$ transition operators.

%-----------------------------------------------------------
%
%     A=150 NME parameter variation
%
%-----------------------------------------------------------
\begin{figure}[ht]
\begin{center}
\includegraphics[width=\linewidth]{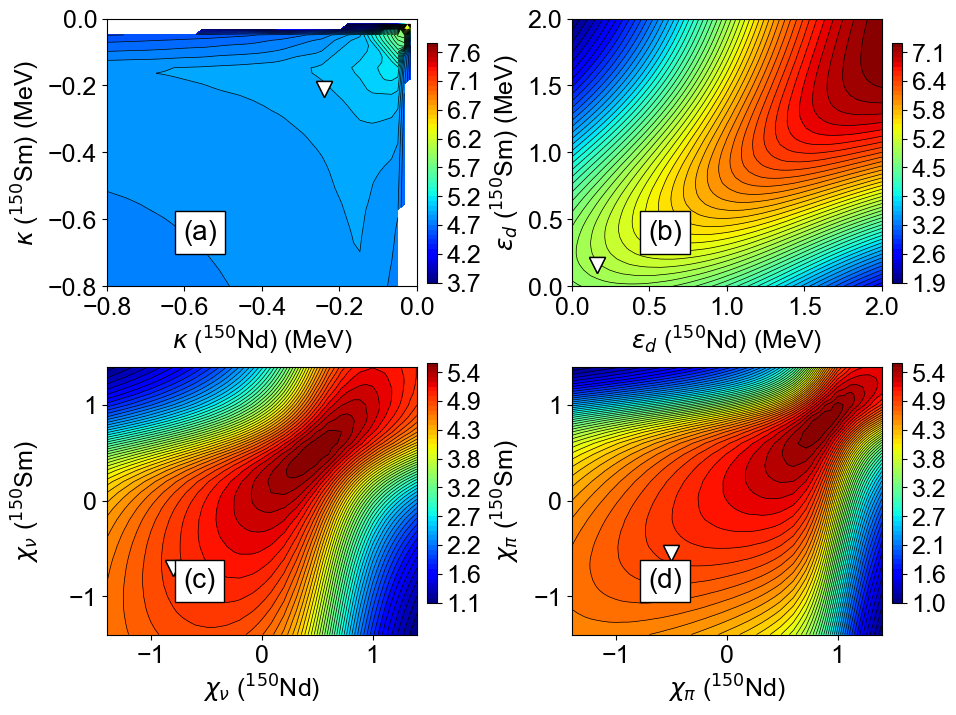}
\caption{
Same as the caption to Fig.~\ref{fig:para-76gese}, but
for the NMEs for the $\znbb$ decay
$^{150}$Nd$(0^+_1)$ $\to$ $^{150}$Sm$(0^+_1)$.
}
\label{fig:para-150ndsm}
\end{center}
\end{figure}

\subsection{IBM-2 Hamiltonian parameters\label{sec:para}}

Even though the IBM-2 Hamiltonian parameters 
are specified by the mapping procedure, 
it is of interest to analyze dependencies 
of the calculated NMEs on these parameters. 
As an illustrative example,
Fig.~\ref{fig:para-76gese} shows contour plots 
of the NMEs for the decay 
$^{76}$Ge$(0^+_1)$ $\to$ $^{76}$Se$(0^+_1)$
in terms of the IBM-2 Hamiltonian parameters 
for the parent and daughter nuclei. 
In this analysis,
only one of the parameters for each even-even nucleus 
is varied, keeping all the other parameters unchanged, 
and the cubic term is not considered for simplicity.

The quadrupole-quadrupole strength 
$\kappa$ is expected to influence significantly 
the spectroscopic properties of each nucleus 
and the NMEs, 
since the term $\hat Q_{\nu} \cdot \hat Q_{\pi}$
is most responsible for determining 
the $d$-boson contents in the wave functions for 
the ground and excited $0^+$ states. 
The relevance of the quadrupole-quadrupole 
strength $\kappa$ was investigated in the 
studies of the $\tnbb$ decays \cite{nomura2024bb} 
and single-$\beta$ decay properties of 
the neutron-rich Zr isotopes \cite{homma2024}.
It was shown in \cite{nomura2024bb} that 
the decrease in magnitude of this parameter led to 
the enhancement of the $\tnbb$-decay NMEs
\cite{nomura2024bb}. 
The parameter sensitivity analysis for 
the mapped IBM-2 in Ref.~\cite{homma2024} 
suggested that by decreasing the magnitude 
$|\kappa|$ the calculated $\beta$-decay 
$\log(ft)$ values of the neutron-rich 
Zr isotopes became larger and consistent 
with data \cite{homma2024}. 
With the decreases in magnitude of the 
quadrupole-quadrupole strengths $\kappa$ for the 
parent ($^{76}$Ge) and daughter ($^{76}$Se) nuclei, 
the NME becomes larger [Fig.~\ref{fig:para-76gese}(a)]. 
These behaviors of the NME are explained by 
the fact that the $d$-boson contributions 
are suppressed by the decreases in the
magnitude $|\kappa|$, while the dominant, 
monopole components in the NME are enhanced.

As shown in Fig.~\ref{fig:para-76gese}(b), 
larger values of the NMEs are obtained 
by increasing the single $d$-boson 
energies $\epsilon_d$, since the monopole 
contributions become even more dominant over 
the quadrupole ones. 
The NMEs appear to be less sensitive to 
the changes in the parameters $\chi_{\nu}$ 
[Fig.~\ref{fig:para-76gese}(c)] and $\chi_{\pi}$ 
[Fig.~\ref{fig:para-76gese}(d)] than to $\kappa$ and $\epsilon_d$. 
In Fig.~\ref{fig:para-76gese}(c),
the largest NMEs are obtained if the values
$\chi_{\nu}\approx$ 0.5 and $-0.5$ are taken 
for $^{76}$Ge and $^{76}$Se, respectively. 
These $\chi_{\nu}$ values are opposite in sign
but are of the same order of magnitude as the
$\chi_{\pi}$ parameter values determined 
by the mapping, that is, $\chi_{\pi}=-0.5$ and 0.5 
for $^{76}$Ge and $^{76}$Se, respectively. 
The sum $\chi_{\nu}+\chi_{\pi}$ that nearly vanishes 
indicates that the quadrupole deformation 
is significantly suppressed, since assuming 
that the quadrupole operator for the total boson system 
is approximately given 
as $\hat Q_{\nu} + \hat Q_{\pi}$ the matrix element 
of the term 
$\chi_{\nu} (d^\+_{\nu} \times \tilde d_{\nu})^{(2)}
+\chi_{\pi} (d^\+_{\pi} \times \tilde d_{\pi})^{(2)}$ 
is significantly reduced. 
The monopole components are, however, supposed to 
play an even more significant role and produce 
the enhanced NMEs, with the above combination 
of the $\chi_{\nu}$ and $\chi_\pi$ values. 
Also in Fig.~\ref{fig:para-76gese}(d) 
the $\chi_{\pi}$ values of 
$\chi_{\pi}\approx$ 0.9 ($^{76}$Ge) and $-0.9$ ($^{76}$Se)
give the largest NMEs, and these values are 
of the same order of magnitude as 
the derived $\chi_{\nu}$ values, 
$-0.9$ ($^{76}$Ge) and 0.9 ($^{76}$Se).

Similar parameter sensitivity analysis
is made for the decay
$^{150}$Nd$(0^+_1)$ $\to$ $^{150}$Sm$(0^+_1)$,
the NMEs of which are predicted to be anomalously large
in the mapped IBM-2 (see Fig.~\ref{fig:nme}).
Variation of the corresponding NME with
parameters $\kappa$, $\epsilon_d$, $\chi_{\nu}$,
and $\chi_{\pi}$ for the parent and daughter
nuclei is shown in Fig.~\ref{fig:para-150ndsm}.
The predicted NME here exhibits only a weak
dependence on the quadrupole-quadrupole
strength $\kappa$, but is rather sensitive
to the single $d$-boson energy $\epsilon_d$.
It appears from Fig.~\ref{fig:para-150ndsm}(b) that,
to reduce the NME value, the $\epsilon_d$ value
either for the parent or daughter nucleus
would need to be increased, with
the other being slightly decreased, so that the
$d$-boson contributions to the NME,
which are positive in sign (see Fig.~\ref{fig:ssdd}),
are suppressed.
The NME for the $^{150}$Nd is sensitive also
to the parameters
$\chi_{\nu}$ and $\chi_{\pi}$.
A large positive value of these parameters
for either the parent or daughter nucleus
would reduce the NME, but is unrealistic
since both nuclei are prolate deformed,
for which large negative values should
be chosen for the $\chi_{\nu}$ and $\chi_{\pi}$
parameters in order to reproduce the
SCMF PESs.

\subsection{Form of the IBM-2 Hamiltonian\label{sec:form}}

Even though the IBM-2 Hamiltonian of \eqref{eq:hb}
(up to the cubic term) has been frequently used
in the previous IBM-2 calculations \cite{IBM},
it is of a simplified form of a more general Hamiltonian,
and one can in principle include some additional boson terms
that may affect the NMEs.
In particular, in the IBM-2 wave functions for the ground state
there may be states with the neutron-proton
mixed symmetry, which is characterized by the
$F$-spin quantum number $F$ that is less than
that for the fully-symmetric states with $F_{\rm max}$,
i.e., $F<F_{\rm max}$.
To exclude the mixed symmetry states from near
the ground state, it has often been considered to
include the so-called Majorana terms in the IBM-2
Hamiltonian \cite{OAIT,IBM}, and these terms were also
shown to have impacts on the $\znbb$-decay NMEs in some
rare-earth nuclei \cite{beller2013}.
In the present framework, however,
the Majorana terms do not appear in the
IBM-2 PES if the equal neutron and proton
deformations, that is,
$\beta_{\nu}=\beta_{\pi}$ and $\gamma_{\nu}=\gamma_{\pi}$,
are assumed.
Only the procedure of mapping the PES is therefore insufficient
to determine the strength parameters for the Majorana
interactions, and it is necessary to develop
an alternative way of deriving the Majorana parameters
from the SCMF calculations.
This would require major extension of the theoretical
framework and is beyond the scope of the
present work.

%-----------------------------------------------------------
%
%     A=76 and 150 spectra with modified params
%
%-----------------------------------------------------------
\begin{figure}[ht]
\begin{center}
\includegraphics[width=\linewidth]{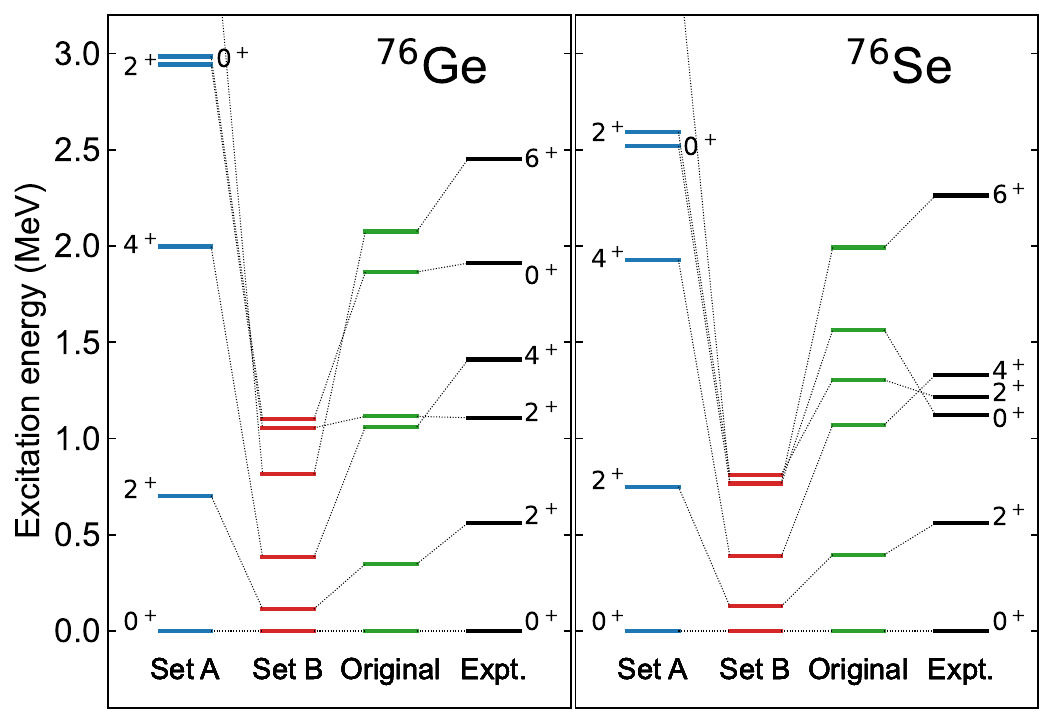}
\includegraphics[width=\linewidth]{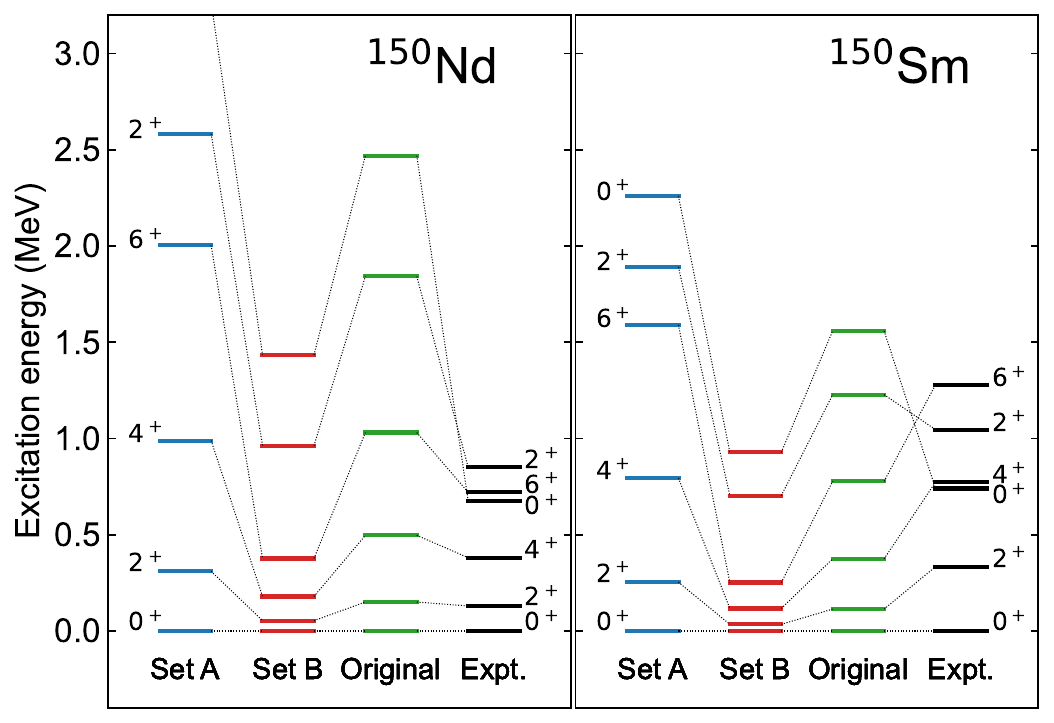}
\caption{
Excitation energies of the $^{76}$Ge, $^{76}$Se,
$^{150}$Nd, and $^{150}$Sm nuclei
computed by the
mapped IBM-2 with the parameter set A and set B
(see the main text for details), and
with the original set of parameters
employed in the present study.
The IBM-2 mapping is based on
the constrained RHB calculations,
and the experimental data
are taken from the Brookhaven National Nuclear Data Center
\cite{data}.
}
\label{fig:modpara-level}
\end{center}
\end{figure}

\subsection{Mapping procedure\label{sec:mapping}}

One might ask if there are
any other sets of parameters than those considered
in this work that also reproduce the topology of SCMF PES,
and what impacts these alternative parameter sets
would have on the NMEs.
The mapping is unique \cite{nomura2010}
under the conditions that
it is carried out only in the vicinity of the global
minimum in the SCMF PES, that is, the topology of the
SCMF PES with the excitation energy of up to several
MeV and/or within the range of the $\beta$ deformation
from 0 to the value that is slightly larger than 
the $\beta_{\rm min}$ corresponding to the global minimum.
The reason why the region of PES has to be thus limited
is that the mean-field configurations near the global
minimum are most relevant for the low-energy collective states.
One should not try to reproduce every detail of the PES
that is very far from the global minimum,
since in that region quasiparticle degrees of freedom
come to play a role, which are by construction
not included in the IBM-2 space.
The uniqueness and ambiguity of the IBM-2 mapping
have been addressed in detail in Ref.~\cite{nomura2010},
and it was shown that even though there are other
possible combinations of the IBM-2 parameters,
that give a perfect fit to the SCMF PES,
these parameters are obtained to fit also those regions
of the SCMF PES that should be excluded for
the above-mentioned reason and are
in most cases just unphysical,
e.g., negative $d$ boson energy,
$\chi_{\nu}$ and/or $\chi_{\pi}$
values that are much larger in magnitude than
the SU(3) limit $\pm\sqrt{7}/2$, 
or some parameters being very far from
those used in empirical calculations.
The mapping procedure may, in principle, result in
infinite numbers of optimal IBM-2 parameters
that are in the vicinity of the values employed
in the present study.
These numerous different combinations of parameters
are, however, quite unlikely
to affect the energies and wave functions, hence
hardly alter the conclusion on the final NME predictions.

It would be nevertheless useful to study
some extreme cases, in which additional sets of the IBM-2
mapping calculations are carried out for each
of the parent and daughter nuclei involved in given
$\znbb$ decays: one with a set of the parameters
in which $|\kappa|$ is increased by 50\% (Set A),
and the other in which $|\kappa|$ is decreased by 50\%
(Set B). In both Set A and Set B, other
parameters are readjusted to fit the SCMF PES.
As illustrative examples, the $\znbb$ decays
$^{76}$Ge$(0^+_1)$ $\to$ $^{76}$Se$(0^+_1)$ and
$^{150}$Nd$(0^+_1)$ $\to$ $^{150}$Sm$(0^+_1)$
are studied, with the microscopic input
from the RHB.
The adopted parameter set A and set B for each
nucleus are summarized in Table~\ref{tab:modpara-para}.
The parameter set A is determined so that not only
the topology of the PES near the global minimum
but also the region far from it
should be reproduced, which is, however, considered
irrelevant here.
For determining the parameter set B,
with the decrease in magnitude of the strength
$\kappa$ the $d$-boson energy $\epsilon_d$ also
has to be significantly reduced to be negative
values for $^{150}$Nd and $^{150}$Sm,
which are however unrealistic,
and the $\chi_{\nu}$ and $\chi_{\pi}$ parameters
are changed.

The resultant energy spectra for $^{76}$Ge, $^{76}$Se,
$^{150}$Nd, and $^{150}$Sm are shown in
Fig.~\ref{fig:modpara-level}.
The parameter set A and set B
lead to the energy spectra that are overall
stretched and compressed, respectively, with
respect to those obtained with
the original parameter set.
Table~\ref{tab:modpara-nme} lists
the calculated values of the NMEs with nine different combinations
of the parameter sets used for parent and daughter
nuclei.
In the cases in which the parameter set A (set B) is
considered either for parent or daughter nuclei in
the $^{76}$Ge $\to$ $^{76}$Se decay,
the NME is enhanced (reduced).
The changes in the NME for this decay process
appear to be dominated mainly by the changes
in the $d$-boson energy $\epsilon_d$, since
as seen in Fig.~\ref{fig:para-76gese}(b) the
increase (decrease) of this parameter leads
to a larger (smaller) NME.
Also for the $^{150}$Nd $\to$ $^{150}$Sm decay,
the use of the set A parameters generally
results in larger NMEs than in the case of the
original set of the parameters.
As seen from Table~\ref{tab:modpara-nme}
the maximum and minimum NMEs values differ
by a factor of 1.8 and 1.7,
for the decays $^{76}$Ge $\to$ $^{76}$Se
and $^{150}$Nd $\to$ $^{150}$Sm, respectively.

%-----------------------------------------------------------
%
%     Modified params
%
%-----------------------------------------------------------
\begin{table}
\caption{\label{tab:modpara-para}
Sets of the IBM-2 parameters for $^{76}$Ge, $^{76}$Se,
$^{150}$Nd, and $^{150}$Sm employed in this
study denoted original, and modified parameter sets
denoted set A and set B.
See the main text for details.
The IBM-2 mapping is based on
the constrained RHB calculations.
}
 \begin{center}
 \begin{ruledtabular}
  \begin{tabular}{lcccccc}
 & & $\epsilon_d$ (MeV) & $\kappa$ (MeV) 
 & $\chi_{\nu}$ & $\chi_{\pi}$ & $\theta$ (MeV) \\
\hline
\multirow{3}{*}{$^{76}$Ge} 
& Original & $0.60$ & $-0.38$ & $-0.90$ & $-0.50$ & $0.50$ \\
& Set A & $1.30$ & $-0.57$ & $-0.90$ & $-0.50$ & $0.50$ \\ 
& Set B & $0.07$ & $-0.19$ & $-1.00$ & $-0.95$ & $0.50$ \\ 
[1.0ex]
\multirow{3}{*}{$^{76}$Se}
& Original & $0.96$ & $-0.22$ & $0.90$ & $0.50$ & $0.30$ \\ 
& Set A & $1.67$ & $-0.33$ & $0.90$ & $0.50$ & $0.30$ \\ 
& Set B & $0.35$ & $-0.11$ & $0.90$ & $0.65$ & $0.00$ \\
[1.0ex]
\multirow{3}{*}{$^{150}$Nd}
& Original & $0.16$ & $-0.24$ & $-0.80$ & $-0.50$ & $0.00$ \\ 
& Set A & $0.80$ & $-0.36$ & $-0.70$ & $-0.40$ & $0.00$ \\ 
& Set B & $-0.28$ & $-0.12$ & $-1.05$ & $-1.05$ & $0.00$ \\
[1.0ex]
\multirow{3}{*}{$^{150}$Sm}
& Original & $0.16$ & $-0.21$ & $-0.70$ & $-0.55$ & $0.00$ \\
& Set A & $0.68$ & $-0.315$ & $-0.60$ & $-0.50$ & $0.00$ \\
& Set B & $-0.20$ & $-0.105$ & $-1.02$ & $-0.95$ & $0.00$ \\
  \end{tabular}
 \end{ruledtabular}
 \end{center}
\end{table}

%-----------------------------------------------------------
%
%     NMEs with modified params
%
%-----------------------------------------------------------
\begin{table}
\caption{\label{tab:modpara-nme}
NMEs calculated within the mapped IBM-2
with the sets of parameters denoted original,
set A, and set B (see the main text for details)
for parent and daughter nuclei of the
$\znbb$ decays $^{76}$Ge$(0^+_1)$ $\to$ $^{76}$Se$(0^+_1)$
and $^{150}$Nd$(0^+_1)$ $\to$ $^{150}$Sm$(0^+_1)$.
The IBM-2 mapping is based on
the constrained RHB calculations.
}
 \begin{center}
 \begin{ruledtabular}
  \begin{tabular}{ccccc}
 & & \multicolumn{3}{c}{$^{76}$Ge} \\ \cline{3-5}
 & & Set A & Original & Set B \\
\hline
\multirow{3}{*}{$^{76}$Se} & Set A & $3.205$ & $2.937$ & $2.089$ \\ 
 & Original & $3.032$ & $2.806$ & $2.040$ \\ 
 & Set B & $2.485$ & $2.345$ & $1.772$ \\ 
[1.0ex]
 & & \multicolumn{3}{c}{$^{150}$Nd} \\ \cline{3-5}
 & & Set A & Original & Set B \\
\hline
\multirow{3}{*}{$^{150}$Sm} & Set A & $5.525$ & $4.930$ & $3.227$ \\ 
 & Original & $5.079$ & $4.929$ & $3.771$ \\ 
 & Set B & $3.321$ & $3.777$ & $3.898$ \\ 
  \end{tabular}
 \end{ruledtabular}
 \end{center}
\end{table}

%-----------------------------------------------------------
%
%       SCMF PESs for 76Se and 96Zr
%
%-----------------------------------------------------------
\begin{figure}[ht]
\begin{center}
\includegraphics[width=\linewidth]{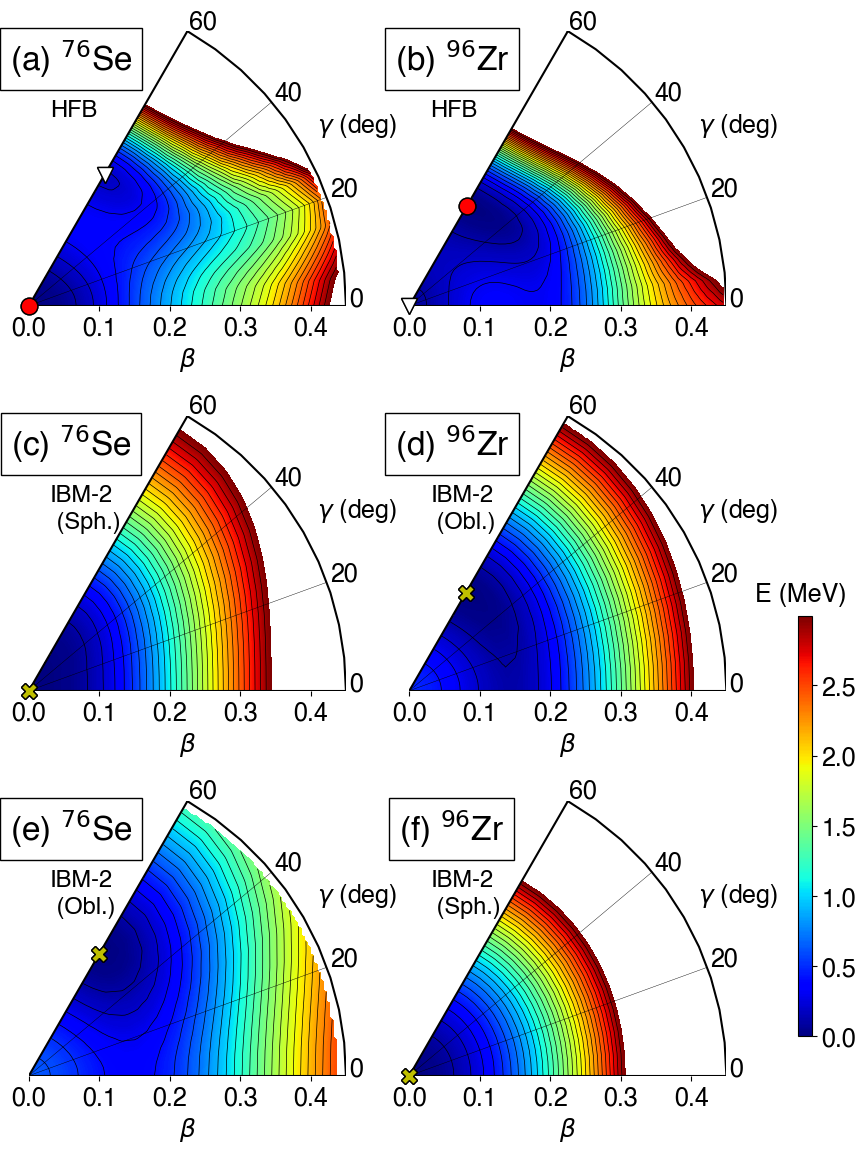}
\caption{
HFB PESs in terms of the triaxial 
quadrupole $(\beta,\gamma)$ deformations for 
(a) $^{76}$Se and (b) $^{96}$Zr, and
the corresponding IBM-2 PESs 
with the Hamiltonian associated with 
the global [(c) and (d)], and 
local [(e) and (f)] minima on the SCMF PESs. 
The global, and local minima on 
the HFB PESs are denoted by the 
solid circle and open triangle, respectively. 
The global minimum in the IBM-2 PESs is
indicated by the cross. 
}
\label{fig:pes-gese}
\end{center}
\end{figure}

\subsection{Coexistence of more than one mean-field minimum\label{sec:co}}

In those nuclei for which the PESs exhibit 
a local minimum close in energy to 
the global minimum, there supposed to be
certain shape mixing, which influences 
the spectroscopic properties and NMEs.
The effects of coexisting minima are 
here analyzed by performing 
two sets of the mapped IBM-2 calculations, 
one in which the Hamiltonian is associated
with the global minimum, and the other in which 
the Hamiltonian is associated with 
a local minimum.

As illustrative cases the HFB mapped IBM-2 
calculations for the decays
$^{76}$Ge$(0^+_1)$ $\to$ $^{76}$Se$(0^+_{1,2})$ and 
$^{96}$Zr$(0^+_1)$ $\to$ $^{96}$Mo$(0^+_{1,2})$ 
are considered. 
The HFB PESs for $^{76}$Se and $^{96}$Zr
are given in Fig.~\ref{fig:pes-gese}, and one 
observes an oblate local minimum
at $\beta\approx0.22$ for $^{76}$Se 
and a spherical local minimum for $^{96}$Zr.
The IBM-2 parameters for the $^{76}$Se
are determined so that the global minimum in the
IBM-2 PES occurs at the same $(\beta,\gamma)$
configuration as in the HFB PES, that corresponds
either to the spherical local or oblate global minimum,
and so that the topology of the HFB PES in the vicinity
of the chosen mean-field minimum should be reproduced.
The resultant parameter values for $^{76}$Se are 
$\epsilon_d=0.8$ (1.0) MeV, $\kappa=-0.2$ ($-0.14$) MeV, 
$\chi_{\nu}=0.4$ (0.4), and $\chi_{\pi}=0.4$ (0.4),
if the energy minimum of the IBM-2 PES is associated
with the spherical local (oblate global) minimum
in the HFB PES.
A similar procedure is applied to $^{96}$Zr,
that is, the mapping is carried out so that
the global minimum in the IBM-2 PES should occur
near the oblate $\beta\approx 0.16,\gamma=60^{\circ}$
(or spherical) configuration, which corresponds to
the global (local) minimum in the HFB PES.
The derived parameters for $^{96}$Zr are
$\epsilon_d=1.8$ (1.24) MeV, $\kappa=-0.18$ ($-0.25$) MeV, 
$\chi_{\nu}=-0.25$ ($-0.25$), and $\chi_{\pi}=0.47$ (0.47),
if the energy minimum of the IBM-2 PES is associated
with the oblate local (spherical global) minimum
in the HFB PES.

\begin{table}
\caption{\label{tab:co-1}
GT, Fermi, tensor, and final $\mzn$ nuclear 
matrix elements for
$^{76}$Ge$(0^+_1)$ $\to$ $^{76}$Se$(0^+_{1})$ and 
$^{96}$Zr$(0^+_1)$ $\to$ $^{96}$Mo$(0^+_{1})$ $\znbb$ decays 
calculated by the HFB-mapped IBM-2, 
with the Hamiltonian associated with the 
global and local minima. 
The mean-field minima are found 
on spherical (``Sph.'') and oblate (``Obl.'') 
configurations, and that configuration 
corresponding to the global minimum is shown in bold.
}
 \begin{center}
 \begin{ruledtabular}
  \begin{tabular}{lccccc}
Decay & config. & $\mgt$ & $\mfe$ & $\mte$ & $\mzn$ \\
\hline
\multirow{2}{*}{$^{76}$Ge $\to$ $^{76}$Se} 
& {\bf Sph.} ($^{76}$Se) & $4.066$ & $-0.163$ & $-0.123$ & $4.045$ \\
& Obl. ($^{76}$Se) & $3.631$ & $-0.144$ & $-0.096$ & $3.625$ \\
[1.0ex]
\multirow{2}{*}{$^{96}$Zr $\to$ $^{96}$Mo} 
& Sph. ($^{96}$Zr) & $4.303$ & $-0.687$ & $0.156$ & $4.885$ \\
& {\bf Obl.} ($^{96}$Zr) & $3.636$ & $-0.558$ & $0.141$ & $4.123$ \\
  \end{tabular}
 \end{ruledtabular}
 \end{center}
\end{table}

\begin{table}
\caption{\label{tab:co-2}
Same as the caption to Table~\ref{tab:co-1}, but 
for the $0^+_1$ $\to$ $0^+_2$ decays. 
}
 \begin{center}
 \begin{ruledtabular}
  \begin{tabular}{lccccc}
Decay & config. & $\mgt$ & $\mfe$ & $\mte$ & $\mzn$ \\
\hline
\multirow{2}{*}{$^{76}$Ge $\to$ $^{76}$Se} 
& {\bf Sph.} ($^{76}$Se) & $0.314$ & $-0.016$ & $-0.037$ & $0.286$ \\
& Obl. ($^{76}$Se) & $1.833$ & $-0.077$ & $-0.078$ & $1.802$ \\
[1.0ex]
\multirow{2}{*}{$^{96}$Zr $\to$ $^{96}$Mo} 
& Sph. ($^{96}$Zr) & $0.781$ & $-0.126$ & $0.028$ & $0.887$ \\
& {\bf Obl.} ($^{96}$Zr) & $0.347$ & $-0.052$ & $0.014$ & $0.393$ \\
  \end{tabular}
 \end{ruledtabular}
 \end{center}
\end{table}

\begin{table}
\caption{\label{tab:ssdd-co}
Monopole ($J=0$) and quadrupole ($J=2$) 
parts of the GT matrix elements $\mgt$ 
for the $\znbb$ decays
$^{76}$Ge$(0^+_1)$ $\to$ $^{76}$Se$(0^+_{1,2})$ and 
$^{96}$Zr$(0^+_1)$ $\to$ $^{96}$Mo$(0^+_{1,2})$. 
That configuration corresponding to the global
minimum on the SCMF PES is indicated in bold.
The IBM-2 mapping is based on
the HFB-SCMF calculations.
}
 \begin{center}
 \begin{ruledtabular}
  \begin{tabular}{lccccc}
\multirow{2}{*}{Decay} & \multirow{2}{*}{Config.} & 
\multicolumn{2}{c}{$0^+_1 \to 0^+_1$} &
\multicolumn{2}{c}{$0^+_1 \to 0^+_2$} \\
\cline{3-4}
\cline{5-6}
 & & $J=0$ & $J=2$ 
& $J=0$ & $J=2$ \\
\hline
\multirow{2}{*}{$^{76}$Ge $\to$ $^{76}$Se} 
& {\bf Sph.} ($^{76}$Se) & $4.230$ & $-0.164$ & $0.101$ & $0.212$  \\
& Obl. ($^{76}$Se) & $3.888$ & $-0.257$ & $1.726$ & $0.107$ \\
\multirow{2}{*}{$^{96}$Zr $\to$ $^{96}$Mo} 
& Sph. ($^{96}$Zr) & $4.529$ & $-0.226$ & $0.814$ & $-0.033$ \\
& {\bf Obl.} ($^{96}$Zr) & $4.032$ & $-0.396$ & $-0.398$ & $0.051$ \\
  \end{tabular}
 \end{ruledtabular}
 \end{center}
\end{table}

The resultant NMEs are given in Table~\ref{tab:co-1} 
and Table~\ref{tab:co-2} 
for the $0^+_1$ $\to$ $0^+_1$ 
and $0^+_1$ $\to$ $0^+_2$ decays, 
respectively. For the $^{76}$Ge $\to$ $^{76}$Se decay, 
the IBM-2 mapping calculation based on the oblate 
local minimum in $^{76}$Se provides the GT, Fermi and tensor 
matrix elements for the $0^+_1 \to 0^+_1$ decay 
that are smaller in magnitude than
those based on the spherical global minimum
and gives the final NME that is
smaller than that obtained with
the spherical configuration
by approximately 23 \%
(see Table~\ref{tab:co-1}). 
Also for the $^{96}$Zr$(0^+_1)\to^{96}$Mo$(0^+_{1})$ decay, 
the mapped IBM-2 based on the spherical local
minimum in $^{96}$Zr, 
gives a larger NME than the calculation 
based on the deformed oblate global minimum.

As seen in Table~\ref{tab:co-2}, 
the NMEs for the $0^+_1 \to 0^+_2$ decays 
depend strongly on whether the mapping is 
carried out at the spherical or deformed 
mean-field minimum. 
Indeed, the mapped IBM-2 NME for the 
$^{76}$Ge$(0^+_1)\to^{76}$Se$(0^+_{2})$ decay, 
calculated by using the deformed oblate 
local minimum, is larger than that obtained from the 
calculation based on the spherical global minimum
by a factor of approximately 6.
However, the calculated 
$^{96}$Zr$(0^+_1)\to^{96}$Mo$(0^+_{2})$ $\znbb$ 
decay NME with the oblate deformed configuration 
is smaller than that with the spherical
local minimum by a factor of approximately 2.

Table~\ref{tab:ssdd-co} gives 
monopole ($J=0$) and quadrupole ($J=2$) 
components of the GT matrix elements, 
calculated by the IBM-2 
corresponding to the global and local minima.
For the $0^+_1$ $\to$ $0^+_1$ $\znbb$ decays of 
both $^{76}$Ge and $^{96}$Zr,
the monopole contributions
to $\mgt$
resulting from the spherical 
configuration are larger in magnitude than those 
from the oblate deformed configuration, while 
the quadrupole contributions become minor if 
the mapping is made at the spherical configuration. 
The interpretation of the results 
for the $0^+_1$ $\to$ $0^+_2$ decay of $^{76}$Ge
is not straightforward,
since the calculation based on the oblate deformed 
configuration results in a larger monopole 
GT matrix elements than that based 
on the spherical configuration.
This finding indicates that significant degrees of 
shape mixing are supposed to enter 
the IBM-2 $0^+_2$ wave functions for $^{76}$Se. 
To explicitly take into account the 
coexisting minima and their mixing, the IBM-2 should 
be extended to include intruder configurations 
and their couplings 
with the normal configuration \cite{duval1981}. 
This extension would require a major modification 
of the present theoretical framework, and 
is beyond the scope of this study.

%-----------------------------------------------------------
%
%     NMEs SDI vs EDF
%
%-----------------------------------------------------------
\begin{figure}[ht]
\begin{center}
\includegraphics[width=\linewidth]{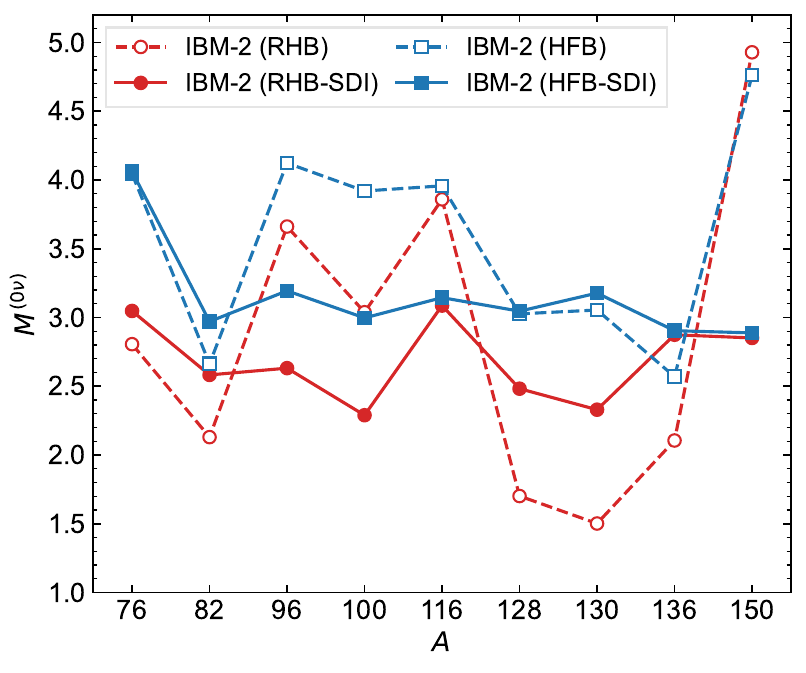}
\caption{NMEs
computed with the pair structure constants 
determined by the inputs from the RHB and
HFB SCMF calculations, and
with those obtained from the surface-delta 
interactions (SDIs) of Ref.~\cite{barea2009}.}
\label{fig:nme-sdi}
\end{center}
\end{figure}

%-----------------------------------------------------------
%
%     NMEs SDI vs EDF
%
%-----------------------------------------------------------
\begin{figure}[ht]
\begin{center}
\includegraphics[width=\linewidth]{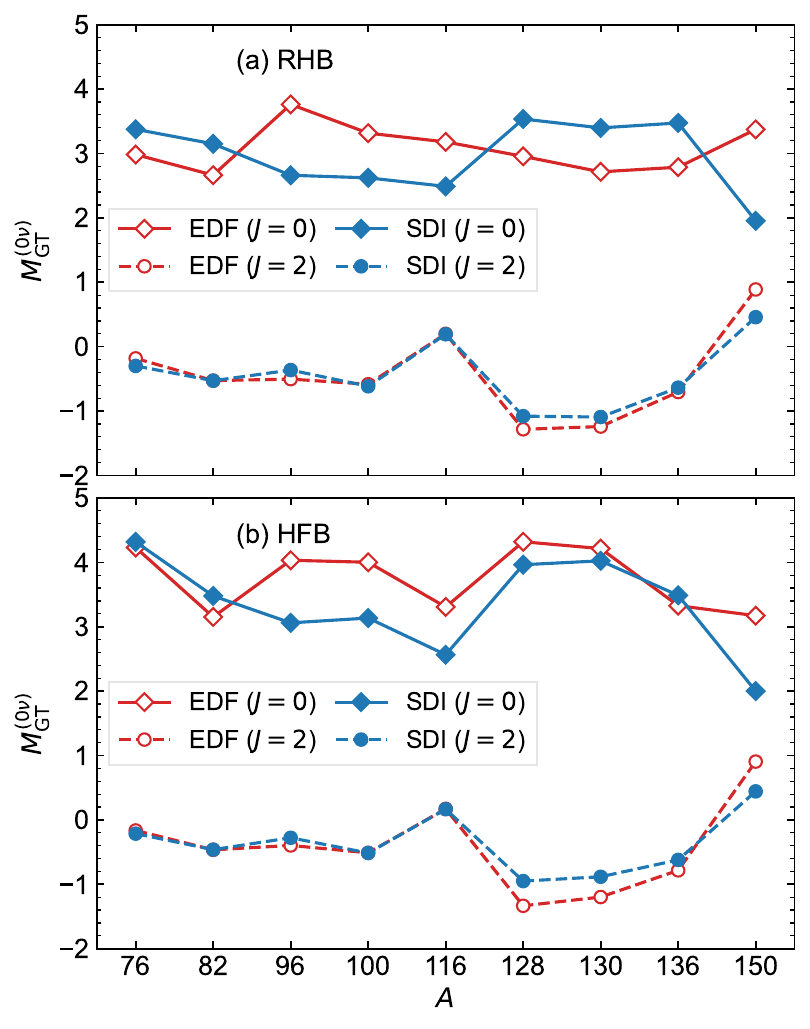}
\caption{Decomposition of the $\mgt$ 
into monopole ($J=0$) and quadrupole ($J=2$) 
components, 
obtained by using the pair structure 
constants resulting from the SCMF and SDI inputs.}
\label{fig:sdi-ssdd}
\end{center}
\end{figure}

\subsection{Pair structure constants\label{sec:sdi}}

The $\znbb$ NME predictions should  
depend on the coefficients 
$\alpha_j$ and $\beta_{j_1j_2}$ in the
pair creation operators [Eqs.~(\ref{eq:spair}) 
and (\ref{eq:dpair})], which also appear
in the coefficients of the 
$\beta\beta$ operators, $A_{\rho}(j)$
[(\ref{eq:ajr-0}) and (\ref{eq:ajr-1})]
and $B_{\rho}(j_1j_2)$ [(\ref{eq:bjr-0}) and (\ref{eq:bjr-1})].
In the earlier IBM-2 NME calculations 
\cite{barea2009,barea2015}, 
the pair structure constants were obtained from 
the diagonalization of the 
shell-model Hamiltonian employing 
surface delta interactions (SDIs), 
and the relative sign of $\alpha_j$ to $\beta_{j_1j_2}$ 
was determined using the formula of Eq.~(\ref{eq:beta}). 
In the present formalism, 
these parameters are determined in a different way, 
that is, $\alpha_j$ values are calculated for 
each decay process by using the occupation probabilities 
$v_j^2$ [see Eq.~(\ref{eq:alpha})] 
computed by the EDF self-consistent
methods,
and $\beta_{j_1j_2}$ values are calculated 
by using the formula of (\ref{eq:beta}) 
(see also the descriptions in Sec.~\ref{sec:znbb}).

Figure~\ref{fig:nme-sdi} shows 
the NMEs calculated by using the 
SCMF- and SDI-derived pair structure constants. 
Here the SDIs refer to those shown from 
Table XIV to Table XVI of Ref.~\cite{barea2009}, 
which are denoted as ``Set I'', corresponding to 
different neutron and proton major shells. 
The $\znbb$-decay NMEs calculated using the SDIs 
for the decays of $^{76}$Ge, $^{82}$Se, $^{128}$Te, $^{130}$Te, 
and $^{136}$Xe are larger than those calculated 
with the SCMF inputs.
However, the SDI-based NMEs for the $\znbb$ decays 
of $^{96}$Zr, $^{100}$Zr, $^{116}$Cd, and $^{150}$Nd are
smaller than those based on the SCMF calculations.

Figure~\ref{fig:sdi-ssdd} depicts 
decomposition of the $\mgt$ matrix elements 
into monopole and quadrupole components.
The monopole parts of $\mgt$
calculated by employing the SCMF-based pair structure 
constants substantially differ from those
resulting from the SDI-derived pair structure 
constants, while the quadrupole parts are
not sensitive to the values of these constants, 
except for the $^{128}$Te, $^{130}$Te,
and $^{150}$Nd decays in the case of the
RHB input.
The monopole components of the NMEs 
depend only on the $\alpha_j$ 
parameters [see (\ref{eq:ajr})], 
and therefore the following discussion 
concerns sensitivity of the NME results 
to the $\alpha_j$ values.

As illustrative examples, 
Tables~\ref{tab:sdi-texe130} 
and \ref{tab:sdi-ndsm} give the 
neutron and proton pair structure 
constants $\alpha_{j}$ used to compute 
the NMEs of the decays 
$^{130}$Te $\to$ $^{130}$Xe 
and $^{150}$Nd $\to$ $^{150}$Sm, respectively. 
As shown in Fig.~\ref{fig:sdi-ssdd}, 
the $\mgt$ values for the $^{130}$Te ($^{150}$Nd)
decay obtained from the SCMF inputs are 
smaller (larger) than that calculated by 
using the pair structure constants 
computed using the SDIs. 
Tables~\ref{tab:sdi-texe130} 
and \ref{tab:sdi-ndsm} also show the 
quantity in percent
\begin{eqnarray}
\label{eq:alph-diff}
 \Delta\alpha_j = \frac{\alpha_j^{\rm (SCMF)}-\alpha_j^{\rm (SDI)}}{\alpha_j^{\rm (SDI)}} \; ,
\end{eqnarray}
which measures the difference between 
the SCMF-based $\alpha_j^{\rm (SCMF)}$
and SDI-based $\alpha_j^{\rm (SDI)}$ values. 
One sees in Table~\ref{tab:sdi-texe130} that,
for the $^{130}$Te decay, in the case of the 
RHB input $\alpha_j^{\rm (SCMF)}$ values are
in most cases smaller than the $\alpha_j^{\rm (SDI)}$, 
i.e., $\Delta\alpha_j<0$.
The $\alpha_j^{\rm (SCMF)}$ values from the
HFB input appear to be, however, more or less 
close to the SDI counterparts.
As one can see in Table~\ref{tab:sdi-ndsm}, 
the $\alpha_j^{\rm (SCMF)}$ values with both the 
RHB and HFB inputs are, in most cases,
considerably larger than the $\alpha_j^{\rm (SDI)}$ 
values. The $\nu\alpha_{9/2}$ value for the 
neutron $1h_{9/2}$ orbit 
obtained from the RHB is, in particular,
larger than that based on the SDI by more than 
a factor of 4.
These quantitative differences between 
$\alpha_j^{\rm (SCMF)}$ and $\alpha_j^{\rm (SDI)}$ 
appear to be a major source of the deviations
of $\mgt$ with the SCMF input from those obtained from
the SDI, as demonstrated in
Figs.~\ref{fig:nme-sdi} and \ref{fig:sdi-ssdd}.

\begin{table}
\caption{\label{tab:sdi-texe130}
Pair structure constants $\alpha_j$ 
for the neutron and proton 
$3s_{1/2}$, $2d_{3/2}$, $2d_{5/2}$, $1g_{7/2}$, 
and $1h_{11/2}$ orbits employed 
for the operators in (\ref{eq:mzn-1}) for the 
$\znbb$ decay of $^{130}$Te. 
Those $\alpha_j$ values determined by the 
occupation probabilities $v^2_j$ computed with 
the RHB and HFB SCMF methods,
and those derived from the SDI are shown. 
The quantity $\Delta\alpha_j$ (in percent) is  
defined in Eq.~(\ref{eq:alph-diff}). 
The $\alpha_j$ values based on the SDI 
for neutrons and protons are taken from Table~XV
of Ref.~\cite{barea2009}, in which they 
are denoted ``Neutrons (I) (holes)'' and 
``Protons (particles).''
}
 \begin{center}
 \begin{ruledtabular}
  \begin{tabular}{lccccc}
 \multirow{2}{*}{Orbit} & SDI & \multicolumn{2}{c}{RHB} & \multicolumn{2}{c}{HFB} \\
\cline{3-4}\cline{5-6}
 & $\alpha_j$ & $\alpha_j$ & $\Delta\alpha_j$ (\%) & $\alpha_j$ & $\Delta\alpha_j$ (\%) \\
\hline
$\nu s_{1/2}$ & $-0.999$ & $-0.999$ & $0$ & $-0.830$ & $-17$ \\
$\nu d_{3/2}$ & $-1.395$ & $-0.847$ & $-39$ & $-0.927$ & $-34$ \\
$\nu d_{5/2}$ & $-0.469$ & $-0.411$ & $-12$ & $-0.405$ & $-14$ \\
$\nu g_{7/2}$ & $-0.357$ & $-0.316$ & $-11$ & $-0.453$ & $27$ \\
$\nu h_{11/2}$ & $1.287$ & $1.453$ & $13$ & $1.431$ & $11$ \\
%---
$\pi s_{1/2}$ & $0.382$ & $0.226$ & $-41$ & $0.379$ & $-1$ \\
$\pi d_{3/2}$ & $0.414$ & $0.320$ & $-23$ & $0.509$ & $23$ \\
$\pi d_{5/2}$ & $0.817$ & $0.581$ & $-29$ & $1.154$ & $41$ \\
$\pi g_{7/2}$ & $1.769$ & $1.879$ & $6$ & $1.602$ & $-9$ \\
$\pi h_{11/2}$ & $-0.406$ & $-0.320$ & $-21$ & $-0.424$ & $4$ \\
  \end{tabular}
 \end{ruledtabular}
 \end{center}
\end{table}

\begin{table}
\caption{\label{tab:sdi-ndsm}
Pair structure constants $\alpha_j$ for
the neutron $3p_{1/2}$, $3p_{3/2}$, 
$2f_{5/2}$, $2f_{7/2}$, $1h_{9/2}$, 
and $1i_{13/2}$ orbits, and proton 
$3s_{1/2}$, $2d_{3/2}$, $2d_{5/2}$, $1g_{7/2}$, 
and $1h_{11/2}$ orbits used for the $^{150}$Nd decay. 
The $\alpha_j$ values based on the SDI for neutrons 
are taken from Table~XVI 
of Ref.~\cite{barea2009}, in which they 
are denoted ``Neutrons (I)'' and those 
values for protons are adopted from 
Table~XV in the same reference, denoted 
as ``Protons.''
}
 \begin{center}
 \begin{ruledtabular}
  \begin{tabular}{lccccc}
 \multirow{2}{*}{Orbit} & SDI & \multicolumn{2}{c}{RHB} & \multicolumn{2}{c}{HFB} \\
\cline{3-4}\cline{5-6}
 & $\alpha_j$ & $\alpha_j$ & $\Delta\alpha_j$ (\%) & $\alpha_j$ & $\Delta\alpha_j$ (\%) \\
\hline
$\nu p_{1/2}$ & $-0.418$ & $-0.275$ & $-34$ & $-0.373$ & $-11$ \\
$\nu p_{3/2}$ & $-0.572$ & $-0.318$ & $-44$ & $-0.507$ & $-11$ \\
$\nu f_{5/2}$ & $-0.371$ & $-0.451$ & $22$ & $-0.509$ & $37$ \\
$\nu f_{7/2}$ & $-2.188$ & $-1.070$ & $-51$ & $-1.752$ & $-20$ \\
$\nu h_{9/2}$ & $-0.390$ & $-1.752$ & $349$ & $-1.141$ & $192$ \\
$\nu i_{13/2}$ & $0.349$ & $0.411$ & $18$ & $0.506$ & $45$ \\
%---
$\pi s_{1/2}$ & $0.382$ & $0.506$ & $32$ & $0.627$ & $64$ \\
$\pi d_{3/2}$ & $0.414$ & $0.789$ & $90$ & $0.779$ & $88$ \\
$\pi d_{5/2}$ & $0.817$ & $1.804$ & $121$ & $1.505$ & $84$ \\
$\pi g_{7/2}$ & $1.769$ & $0.603$ & $-66$ & $1.119$ & $-37$ \\
$\pi h_{11/2}$ & $-0.406$ & $-0.740$ & $82$ & $-0.657$ & $62$ \\
  \end{tabular}
 \end{ruledtabular}
 \end{center}
\end{table}

\section{Summary and conclusion\label{sec:summary}}

Summarizing, the present article has proposed a
method of calculating the $\znbb$-decay NME within
the IBM-2 that is based on the nuclear EDF theory.
The Hamiltonian parameters of the IBM-2
providing low-energy spectra
and transition properties of probable $\znbb$ emitting
nuclei and final-state nuclei are determined
by mapping the SCMF deformation energy surface
onto the equivalent IBM-2 energy surface.
The $\znbb$ operators are formulated
within the generalized seniority scheme,
and the NMEs are computed by following 
the steps taken in Ref.~\cite{barea2009}, while
the pair structure constants
are here specified for each decay process by
using the occupation probabilities obtained
from the SCMF calculations.

The calculated low-energy spectra,
$B(E2)$ values for the ground-state yrast
bands, and electric quadrupole and magnetic
dipole moments for the relevant even-even nuclei
have been shown to be, in most cases,
consistent with the experimental data,
regardless of whether the relativistic or nonrelativistic
framework is employed as the input.
This indicates that the nuclear wave functions
computed with the present method are overall considered
to be reliable, given the fact that the IBM-2
parameters are here obtained by
using the microscopic inputs from the SCMF calculations
and that no phenomenological adjustment is made
as in the conventional IBM-2 applications to the
$\znbb$ decay
\cite{barea2009,barea2013,barea2015,deppisch2020}.

Final $\znbb$-decay NMEs
obtained from the
mapped IBM-2 have been shown to fall
into a spectrum of the previously
predicted NME values in other many-body methods,
namely, the QRPA, NSM, IBM-2, EDF(-GCM), IMSRG,
CC, and EFT approaches.
In seven out of the ten considered decay processes,
the mapped IBM-2 gives smaller NMEs than those
of the previous IBM-2 predictions
\cite{barea2015,deppisch2020}.
These deviations arise naturally from the fact
in the present work the IBM-2 parameters,
and pair structure constants are computed
by using the results of the microscopic SCMF
calculations.
The relatively small NMEs in the mapped IBM-2
are accounted for by the fact that
the monopole components of the NME are
not significant or are canceled by large quadrupole components.
The monopole-quadrupole balance in the NMEs is, in turn,
determined largely by the fact that
the SCMF PESs, upon which the mapping is based,
generally exhibit pronounced deformations.
Also, the mapped IBM-2 with input from the RHB
systematically gives lower NMEs than that from the HFB,
since the RHB PESs for most even-even nuclei
exhibit a sharper potential valley than the HFB PESs.
Compared with alternative approaches,
the present NMEs are overall smaller than
the QRPA, IBM-2, and EDF-GCM ones, but are rather close
to the NSM and more recent {\it ab initio} values.
Using the estimated $\taubb$ with 90\% C.L. and the NMEs,
the lowest upper limit of the neutrino mass
$\braket{m_{\nu}}<0.120$ is provided in the
present study for the $^{76}$Ge $\to$ $^{76}$Se decay.

The comparisons of the calculated excitation energies
and $B(E2)$ transition rates with the experimental data
also suggest current limitations and possible
improvements of the mapped IBM-2 framework
for the $\znbb$-decay predictions.
In particular, the high-lying $2^+_1$ and $4^+_1$
levels of $^{96}$Zr imply the effects of the
$N=56$ subshell closure and shape coexistence,
which are not accounted for by the present framework.
Also, for those nuclei with $A \leqslant 100$
the mapped IBM-2 significantly
overestimates the $0^+_2$ energies, and
underestimates the $B(E2;0^+_2 \to 2^+_1)$
transition strengths.
The deficiencies in describing properties of the excited
$0^+$ states are attributed to the fact that
the present mapped IBM-2 suggests an unexpectedly
large deformation and generates an energy spectrum
with features of the deformed rotor.
A possible remedy is to incorporate the configuration mixing
of different intrinsic shapes, which is expected
to lower the $0^+_2$ energies.

Since the mapping is based on the results of the
SCMF calculations, the IBM-2 parameters determined by the
PES-mapping procedure also reflect the properties
of and conditions in the SCMF methods and/or
the chosen effective interactions, which may
be a possible source of uncertainties in the
NME predictions.
Other sources of uncertainties could be the
parameters and form of the IBM-2 Hamiltonian,
mapping procedure, and inputs to calculate
the pair structure constants.
In particular,
among the considered IBM-2 parameters the single $d$-boson
energies and quadrupole-quadrupole interaction
strength have been shown to play a crucial role
in determining the NME, and these parameters
are sensitive to the topology of the SCMF PES.
The form of the IBM-2 Hamiltonian adopted in the present
study is also not complete, and a more realistic
study would involve other interaction
terms such as the Majorana terms.
The determination of the Majorana interaction strengths
would require certain extensions of the mapping
procedure, which is worked out in a future study.
Furthermore, intermediate states in the adjacent
odd-odd nuclei to the even-even ones may not be
negligible for a precise description of the NME.
An extension of the present framework beyond the
closure approximation would address the roles played
by the intermediate states, which would also serve
to analyze explicitly potential impacts of the
single-particle properties on the NMEs.
Finally, additional collective
degrees of freedom concerning 
the octupole and hexadecapole
shape oscillations, and
the dynamical pairing could affect
the NMEs, and these correlations will
be incorporated
in the IBM mapping procedure.

To conclude, the current study presents a new addition
to the $\znbb$ decay NMEs predictions,
which are rigorously investigated
in the field of low-energy nuclear physics.
On the basis of the microscopic SCMF methods with
the nuclear EDFs,
the present theoretical framework is able to
predict detailed spectroscopic properties
for those nuclei that are far from the stability,
and provides $\znbb$-decay NMEs in principle
for any nuclides including
strongly deformed nuclei in heavy-mass and
open-shell regions in a systematic and computationally
feasible way.
As its initial implementation, 
the mapped IBM-2 approach demonstrates its potential
to study the $\znbb$ decay, and the corresponding
results present a certain step toward an
accurate NME prediction and understanding of
the $\znbb$ decays, which are required for new-generation
experiments and are of crucial importance
in nuclear and other domains of physics.

\appendix

\section{Details about formulas\label{sec:app}}

\subsection{Form factors\label{sec:hoc}}

Terms that appear in the form factors 
$\tilde{h}^{\alpha}(p)$ 
(\ref{eq:fehoc})--(\ref{eq:tehoc})
have the following forms. 
\begin{align}
& \tilde h^{\rm F}_{VV}(p) = \gv^2(p^2) \\
& \tilde h^{\rm GT}_{AA}(p) = \ga^2(p^2) \\
& \tilde h^{\rm GT}_{AP}(p)
= -\frac{2}{3} \ga^2(p^2)
\frac{p^2}{p^2+m^2_\pi} \left( 1-\frac{m_\pi^2}{M_A^2}\right) \\
& \tilde h^{\rm GT}_{PP}(p)
= \ga^2(p^2)\left[
\frac{1}{\sqrt{3}}
\frac{p^2}{p^2+m^2_\pi} 
\left( 1-\frac{m_\pi^2}{M_A^2}\right)
\right]^2 \\
& \tilde h^{\rm GT}_{MM}(p) 
= \gv^2(p^2) \frac{2}{3} 
\frac{\kappa_\beta^2 p^2}{4m_p^2} \\
& \tilde h^{\rm T}_{AP}(p) 
= -3 \tilde h^{\rm GT}_{AP}(p) \\
& \tilde h^{\rm T}_{PP}(p) 
= -3 \tilde h^{\rm GT}_{PP}(p) \\
& \tilde h^{\rm T}_{MM}(p) 
=  \frac{3}{2} \tilde h^{\rm GT}_{MM}(p) \; ,
\end{align}
with $m_\pi=140$ MeV/c$^2$ and $m_p=939$ MeV/c$^2$ 
being the pion and proton masses, respectively, 
and with $\kappa_\beta=3.70$ being the isovector 
anomalous magnetic moment of the nucleon. 
The factors $\gv(p^2)$ and $\ga(p^2)$ 
take into account the finite nucleon size effect, 
and take the forms
\begin{align}
 & \gv(p^2) = \gv \left(1+\frac{p^2}{M_V^2}\right)^{-2} \\
 & \ga(p^2) = \ga \left(1+\frac{p^2}{M_A^2}\right)^{-2} \; ,
\end{align}
with the cutoff 
$M_V^2=0.71$ (GeV/c$^2$)$^2$ \cite{DUMBRAJS1983} 
and $M_A=1.09$ GeV/c$^2$ \cite{Schindler2007}.
Also the extra factor 3 for the
$\tilde h^{\rm T}_{AP}(p)$, $\tilde h^{\rm T}_{PP}(p)$,
and $\tilde h^{\rm T}_{MM}(p)$ terms arises
because of the present definition of tensor operator
$S_{12}=({\boldsymbol{\sigma}}_1\cdot\hat{\boldsymbol{r}}_{12})
({\boldsymbol{\sigma}}_2\cdot\hat{\boldsymbol{r}}_{12})-{\boldsymbol{\sigma}}_1\cdot{\boldsymbol{\sigma}}_{2}/3$,
where $\boldsymbol{r}_{12}=\boldsymbol{r}_{1}-\boldsymbol{r}_{2}$
and $\hat{\boldsymbol{r}}_{12}=\boldsymbol{r}_{12}/|\boldsymbol{r}_{12}|$.

\subsection{Calculation of fermion two-body matrix elements\label{sec:tbme}}

The fermion two-body matrix element
$O_{\alpha}(j_1j_2j_1'j_2';J)$ in 
Eq.~(\ref{eq:oalph-1}) is given as
\begin{align}
& O_{\alpha}(j_1j_2j_1'j_2';J)
\nonumber\\
&= \sum_{k_1=|l_1-l_1'|}^{l_1+l_1'}
\sum_{k_2=|l_2-l_2'|}^{l_2+l_2'}
\sum_{k=k_{\rm min}}^{k_{\rm max}}
i^{k_1-k_2+\lambda}
\hat k_1^2
\hat k_2^2
\nonumber\\
& \times (k_10k_20|\lambda0)(-1)^{s_2+k_1}
\left\{
\begin{array}{ccc}
k_1 & s_1 & k \\
s_2 & k_2 & \lambda
\end{array}
\right\}
\nonumber\\
&\times 
(-1)^{j_2+j_1'+J}
\left\{
\begin{array}{ccc}
j_1 & j_2 & J \\
j_2' & j_1' & k
\end{array}
\right\}
\nonumber\\
& \times
\hat k \hat{\jmath}_1 \hat{\jmath}_1'
\left\{
\begin{array}{ccc}
1/2 & l_1 & j_1 \\
1/2 & l_1' & j_1' \\
s_1 & k_1 & k
\end{array}
\right\}
\hat k \hat{\jmath}_2 \hat{\jmath}_2'
\left\{
\begin{array}{ccc}
1/2 & l_2 & j_2 \\
1/2 & l_2' & j_2' \\
s_2 & k_2 & k
\end{array}
\right\}
\nonumber\\
& \times
\braket{1/2\|\Sigma^{(s_1)}\|1/2}
(-1)^{-k_1} \hat l_1 (l_10k_10|l_1'0)
\nonumber\\
& \times
\braket{1/2\|\Sigma^{(s_2)}\|1/2}
(-1)^{-k_2} \hat l_2 (l_20k_20|l_2'0)
\nonumber \\
& \times
R^{(k_1,k_2,\lambda)}
(n_1,l_1,n_2,l_2, n_1',l_1',n_2',l_2') \; ,
\end{align}
where $\braket{1/2\|\Sigma^{(s)}\|1/2}=\sqrt{2(2s+1)}$, 
$k_{\rm min}={\rm max}(|j_1-j_1'|,|j_2-j_2'|)$, and 
$k_{\rm max}={\rm min}(j_1+j_1',j_2+j_2')$. 
$R^{(k_1,k_2,\lambda)}(n_1,l_1,n_2,l_2, n_1',l_1',n_2',l_2')$ 
are radial integrals, which are calculated 
by the method of Horie and Sasaki \cite{horie1961}:
\begin{align}
\label{eq:rad}
&R^{(k_1,k_2,\lambda)}(n_1,l_1,n_2,l_2, n_1',l_1',n_2',l_2')
\nonumber\\
& = (M_1 M_2)^{-1/2} 
\sum_{s_1=0}^{n_1+n_1'}
\sum_{s_2=0}^{n_2+n_2'}
\nonumber\\
& 
\times \quad
a_{l_1+l_1'+2s_1}(n_1 l_1 , n_1' l_1')
a_{l_2+l_2'+2s_2}(n_2 l_2 , n_2' l_2')
\nonumber\\
& 
\times \quad
f^{(k_1,k_2;\lambda)}
(l_1+l_1'+2s_1, l_2+l_2'+2s_2) \; ,
\end{align}
where $M_i$ ($i=1,2$) is defined by
\begin{align}
 M_i = 2^{n_i+n_i'} n_i! {n_i'}! 
(2l_i+2n_i+1)!!
(2l_i'+2n_i'+1)!! \; ,
\end{align}
\begin{align}
 a_{l+l'+2s}(n l , n' l')
= & (-1)^s \sum_{\mu+\mu'=s}
\left(
\begin{array}{c}
n \\
\mu
\end{array}
\right)
\left(
\begin{array}{c}
n' \\
\mu'
\end{array}
\right)
\nonumber\\
& \times
\frac{(2l+2n+1)!!}{(2l+2\mu+1)!!}
\frac{(2l'+2n'+1)!!}{(2l'+2\mu'+1)!!} \; ,
\end{align}
and 
\begin{align}
& f^{(k_1,k_2;\lambda)}(m_1,m_2)
\nonumber\\
= & \sum_{m=(k_1+k_2)/2}^{(m_1+m_2)/2}
a_{2m} 
\left(
\frac{m_1-k_1}{2}k_1, \frac{m_2-k_2}{2}k_2
\right)
J_m^{(\lambda)}(\nu) \; .
\end{align}
$J_m^{(\lambda)}(\nu)$ are integrals 
\begin{align}
 J_m^{(\lambda)}(\nu) = (2\nu)^{-m}
\int_0^{\infty}
h^{\alpha}(p)
e^{-\frac{p^2}{2\nu}}p^{2m+2}dp \; ,
\end{align}
where $\nu = m_{N}\omega /\hbar$ with 
the nucleon mass $m_{N}$.
Note that, in Ref.~\cite{barea2009},
$h^{\alpha}(p)$ is referred to as $v_{\lambda}(p)$.
If the neutrino potential is given in 
the coordinate representation,
$J_m^{(\lambda)}(\nu)$ are given by the formula
\begin{align}
 & J_m^{(\lambda)}(\nu) 
= \sum_{m}^{m-\frac{\lambda}{2}}
(-1)^{\mu}
\left(
\begin{array}{c}
m-\frac{\lambda}{2} \\
\mu
\end{array}
\right)
\frac{(2m+\lambda+1)!!}{2^m (2\lambda+2\mu+1)!!}
\nonumber\\
& \times
\sqrt{\frac{2}{\pi}} 
\nu^{\frac{\lambda+3}{2}+\mu}
\int_0^{\infty}
H_\alpha(r)
e^{-\frac{\nu r^2}{2}}
r^{\lambda+2\mu+2}dr  \; .
\end{align}
The oscillator parameter $\nu$ is here parameterized 
as $\nu = \nu_0 A^{-1/3}$, with $\nu_0 = 0.994$ fm$^{-2}$ 
and with $A$ being the mass number.

To restore isospin symmetry,
an approximate method introduced in Ref.~\cite{barea2015} 
is adopted, that is, the radial integral 
$R^{(k_1,k_2,\lambda)}(n_1,l_1,n_2,l_2, n_1',l_1',n_2',l_2')$ 
in (\ref{eq:rad}) is replaced with 
\begin{align}
& R^{(k_1,k_2,\lambda)}(n_1,l_1,n_2,l_2, n_1',l_1',n_2',l_2')
\nonumber\\
&-\delta_{k_10}\delta_{k_20} \delta_{k0}
\delta_{\lambda 0} 
\delta_{n_1n_1'} \delta_{j_1j_1'} \delta_{l_1l_1'}
\delta_{n_2n_2'} \delta_{j_2j_2'} \delta_{l_2l_2'}
\nonumber \\
& \times
R^{(0,0,0)}(n_1,l_1,n_2,l_2, n_1',l_1',n_2',l_2') \; ,
\end{align}
so that the Fermi 
transition matrix element for the $\tnbb$ decay 
should vanish 
and that the one for the $\znbb$ decay, $\mfe$, should 
be reduced appreciably.

\subsection{Formulas for the two-boson transfer operators\label{sec:abjr}}

The formulas for the coefficients 
$A_\rho(j)$ and 
$B_\rho(j_1j_2)$ in (\ref{eq:mzn-1}) 
are found in Table.~XVII of Ref.~\cite{barea2009}. 
For like-particle protons and like-hole neutrons, 
\begin{align}
\label{eq:ajr}
 A_\rho({j}) 
= \frac{\sqrt{N_{\rho}+1}(N_{\rho}!)^2}
{\eta_{2N_{\rho},0,0}\eta_{2N_{\rho}+2,0,0}}
\hat{\jmath}
\alpha_{j}
\sum_{s=0}^{N_{\rho}}
\left[
\frac{\alpha_{j}^s 
\eta_{2N_{\rho}-2s,0,0}}{(N_{\rho}-s)!}
\right]^2 \; ,
\end{align}
while for like-hole protons and like-particle neutrons 
the above expression is multiplied by $-1$ and 
$N_\rho$ should be replaced with $N_{\rho}-1$. 
Here 
\begin{align}
 \eta_{2N_{\rho},0,0}^2 = \left(N_{\rho}!\right)^2
\sum_{m_1,\ldots,m_k; \sum{m_i}=N_{\rho}}
\left\{
\Pi_{i=1}^k
\alpha_{j_i}^{2m_i}
\left(
\begin{array}{c}
 \Omega_{j_i} \\
m_i
\end{array}
\right)
\right\} \; .
\end{align}
The $B_\rho(j_1j_2)$ coefficients are
\begin{align}
\label{eq:bjr}
 B_\rho(j_1j_2) = (-1)^{j_1+j_2+1}
\sqrt{1+\delta_{j_1j_2}}
\frac{\eta^2_{2N_{\rho}+2,2,2}(j_1j_2)}
{\eta_{2N_{\rho},0,0}\eta_{2N_{\rho}+2,0,0}}
\beta_{j_1j_2}
\end{align}
for like-particle protons and like-hole neutrons, 
and similar expressions are used for 
like-hole protons and like-particle neutrons, 
with the replacement of $N_{\rho}$ with $N_{\rho}-1$ and 
with the factor $(-1)^{\jr+\jr'}$ omitted. 
Note 
\begin{align}
 \eta^2_{2N_{\rho},2,2} = \sum_{j_1 \leqslant j_2}
\beta_{j_1 j_2}^2 \eta^2_{2N_{\rho},2,2}(j_1j_2)
\end{align}
with
\begin{align}
 \eta^2_{2N_{\rho},2,2}(j_1j_2)
=& \sum_{p=0}^{N_{\rho}-1}
\left[
\frac{(N_{\rho}-1)!}{p!}
\right]^2
(-1)^{N_{\rho}-p-1}
\eta^2_{2p,0,0}
\nonumber\\
&\times \sum_{q=0}^{N_{\rho}-p-1}
\left(
\alpha_{j_1}^{N_{\rho}-p-q-1} \alpha_{j_2}^q
\right)^2 \; .
\end{align}

%-----------------------------------------------------------
%
%       vv_48-82
%
%-----------------------------------------------------------
\begin{table}
\caption{\label{tab:vv_48-82}
Occupation probabilities $v_j^2$ obtained from
the RHB and HFB SCMF calculations used to compute the
pair structure constants [\eqref{eq:alpha} and \eqref{eq:beta}],
corresponding to the single-neutron and single-proton
configurations considered for the $\znbb$ decays
of those nuclei with mass $A=48$, 76, and 82.
}
 \begin{center}
 \begin{ruledtabular}
  \begin{tabular}{lcccccc}
%---- neutron orbits ----
 & & \multicolumn{5}{c}{Neutron orbits} \\
\cline{3-7}
$A$ & Input & $2p_{1/2}$ & $2p_{3/2}$ & $1f_{5/2}$ & $1f_{7/2}$ & $1g_{9/2}$ \\
\hline
\multirow{2}{*}{48} & RHB & $0.005$ & $0.011$ & $0.007$ & $1.000$ & ${}$ \\ 
${}$ & HFB & $0.006$ & $0.016$ & $0.006$ & $1.000$ & ${}$ \\ 
\multirow{2}{*}{76} & RHB & $0.966$ & $0.985$ & $0.982$ & ${}$ & $0.316$ \\ 
${}$ & HFB & $0.945$ & $0.981$ & $0.958$ & ${}$ & $0.335$ \\ 
\multirow{2}{*}{82} & RHB & $0.980$ & $0.990$ & $0.990$ & ${}$ & $0.704$ \\ 
${}$ & HFB & $0.980$ & $0.990$ & $0.984$ & ${}$ & $0.704$ \\ 
[1.0ex]
%---- proton orbits ----
 & & \multicolumn{5}{c}{Proton orbits} \\
\cline{3-7}
$A$ & Input & $2p_{1/2}$ & $2p_{3/2}$ & $1f_{5/2}$ & $1f_{7/2}$ & $1g_{9/2}$ \\
\hline
\multirow{2}{*}{48} & RHB & $0.001$ & $0.002$ & $0.004$ & $0.000$ & ${}$ \\ 
${}$ & HFB & $0.001$ & $0.003$ & $0.003$ & $0.000$ & ${}$ \\ 
\multirow{2}{*}{76} & RHB & $0.116$ & $0.411$ & $0.515$ & ${}$ & $0.017$ \\ 
${}$ & HFB & $0.142$ & $0.530$ & $0.421$ & ${}$ & $0.022$ \\ 
\multirow{2}{*}{82} & RHB & $0.137$ & $0.462$ & $0.793$ & ${}$ & $0.019$ \\ 
${}$ & HFB & $0.199$ & $0.621$ & $0.646$ & ${}$ & $0.034$ \\ 
  \end{tabular}
 \end{ruledtabular}
 \end{center}
\end{table}

%-----------------------------------------------------------
%
%       vv_96-116
%
%-----------------------------------------------------------
\begin{table}
\caption{\label{tab:vv_96-116}
Same as the caption to Table~\ref{tab:vv_48-82},
but for the $A=96$, 100, and 100 nuclei.
}
 \begin{center}
 \begin{ruledtabular}
  \begin{tabular}{lcccccc}
%---- neutron orbits ----
 & & \multicolumn{5}{c}{Neutron orbits} \\
\cline{3-7}
$A$ & Input & $3s_{1/2}$ & $2d_{3/2}$ & $2d_{5/2}$ & $1g_{7/2}$ & $1h_{11/2}$ \\
\hline
\multirow{2}{*}{96} & RHB & $0.091$ & $0.093$ & $0.398$ & $0.254$ & $0.025$ \\ 
${}$ & HFB & $0.094$ & $0.055$ & $0.611$ & $0.108$ & $0.016$ \\ 
\multirow{2}{*}{100} & RHB & $0.095$ & $0.092$ & $0.544$ & $0.381$ & $0.022$ \\ 
${}$ & HFB & $0.157$ & $0.090$ & $0.712$ & $0.230$ & $0.029$ \\ 
\multirow{2}{*}{116} & RHB & $0.434$ & $0.451$ & $0.914$ & $0.950$ & $0.101$ \\ 
${}$ & HFB & $0.538$ & $0.397$ & $0.902$ & $0.807$ & $0.202$ \\ 
[1.0ex]
%---- proton orbits ----
 & & \multicolumn{5}{c}{Proton orbits} \\
\cline{3-7}
$A$ & Input & $2p_{1/2}$ & $2p_{3/2}$ & $1f_{5/2}$ & $1g_{9/2}$ & \\
\hline
\multirow{2}{*}{96} & RHB & $0.859$ & $0.962$ & $0.981$ & $0.151$ & \\ 
${}$ & HFB & $0.784$ & $0.950$ & $0.950$ & $0.188$ & \\ 
\multirow{2}{*}{100} & RHB & $0.909$ & $0.967$ & $0.980$ & $0.336$ & \\ 
${}$ & HFB & $0.896$ & $0.968$ & $0.965$ & $0.345$ & \\ 
\multirow{2}{*}{116} & RHB & $0.988$ & $0.994$ & $0.996$ & $0.901$ & \\ 
${}$ & HFB & $0.991$ & $0.995$ & $0.995$ & $0.897$ & \\
  \end{tabular}
 \end{ruledtabular}
 \end{center}
\end{table}

%-----------------------------------------------------------
%
%       vv_128-136
%
%-----------------------------------------------------------
\begin{table}
\caption{\label{tab:vv_128-136}
Same as the caption to Table~\ref{tab:vv_48-82},
but for the $A=128$, 130, and 136 nuclei.
}
 \begin{center}
 \begin{ruledtabular}
  \begin{tabular}{lcccccc}
%---- neutron orbits ----
 & & \multicolumn{5}{c}{Neutron orbits} \\
\cline{3-7}
$A$ & Input & $3s_{1/2}$ & $2d_{3/2}$ & $2d_{5/2}$ & $1g_{7/2}$ & $1h_{11/2}$ \\
\hline
\multirow{2}{*}{128} & RHB & $0.763$ & $0.826$ & $0.961$ & $0.979$ & $0.537$ \\ 
${}$ & HFB & $0.827$ & $0.782$ & $0.961$ & $0.951$ & $0.556$ \\ 
\multirow{2}{*}{130} & RHB & $0.840$ & $0.885$ & $0.973$ & $0.984$ & $0.662$ \\ 
${}$ & HFB & $0.889$ & $0.861$ & $0.973$ & $0.967$ & $0.669$ \\ 
\multirow{2}{*}{136} & RHB & $0.959$ & $0.975$ & $0.994$ & $0.997$ & $0.933$ \\ 
${}$ & HFB & $0.981$ & $0.978$ & $0.995$ & $0.994$ & $0.928$ \\ 
[1.0ex]
%
%---- proton orbits ----
 & & \multicolumn{5}{c}{Proton orbits} \\
\cline{3-7}
$A$ & Input & $3s_{1/2}$ & $2d_{3/2}$ & $2d_{5/2}$ & $1g_{7/2}$ & $1h_{11/2}$ \\
\hline
\multirow{2}{*}{128} & RHB & $0.006$ & $0.012$ & $0.039$ & $0.339$ & $0.339$ \\ 
${}$ & HFB & $0.015$ & $0.027$ & $0.143$ & $0.247$ & $0.247$ \\ 
\multirow{2}{*}{130} & RHB & $0.005$ & $0.010$ & $0.033$ & $0.345$ & $0.345$ \\ 
${}$ & HFB & $0.014$ & $0.026$ & $0.133$ & $0.255$ & $0.255$ \\ 
\multirow{2}{*}{136} & RHB & $0.006$ & $0.013$ & $0.039$ & $0.583$ & $0.583$ \\ 
${}$ & HFB & $0.023$ & $0.041$ & $0.203$ & $0.423$ & $0.423$ \\
  \end{tabular}
 \end{ruledtabular}
 \end{center}
\end{table}

%-----------------------------------------------------------
%
%       vv_150
%
%-----------------------------------------------------------
\begin{table}
\caption{\label{tab:vv_150}
Same as the caption to Table~\ref{tab:vv_48-82},
but for the $A=150$ nuclei.
}
 \begin{center}
 \begin{ruledtabular}
  \begin{tabular}{lccccccc}
%---- neutron orbits ----
 & & \multicolumn{6}{c}{Neutron orbits} \\
\cline{3-8}
$A$ & Input & $3p_{1/2}$ & $3p_{3/2}$ & $2f_{5/2}$ & $2f_{7/2}$ & $1h_{9/2}$ & $1i_{13/2}$ \\
\hline
\multirow{2}{*}{150} & RHB & $0.012$ & $0.016$ & $0.033$ & $0.186$ & $0.499$ & $0.027$ \\ 
${}$ & HFB & $0.022$ & $0.041$ & $0.041$ & $0.516$ & $0.205$ & $0.040$ \\ 
[1.0ex]
%
%---- proton orbits ----
 & & \multicolumn{6}{c}{Proton orbits} \\
\cline{3-8}
$A$ & Input & $3s_{1/2}$ & $2d_{3/2}$ & $2d_{5/2}$ & $1g_{7/2}$ & $1h_{11/2}$ & \\
\hline
\multirow{2}{*}{150} & RHB & $0.030$ & $0.072$ & $0.376$ & $0.958$ & $0.063$ & \\
${}$ & HFB & $0.075$ & $0.116$ & $0.567$ & $0.760$ & $0.083$ & \\
  \end{tabular}
 \end{ruledtabular}
 \end{center}
\end{table}

\subsection{Single-particle spaces and occupation probabilities\label{sec:spe}}

Tables~\ref{tab:vv_48-82}, \ref{tab:vv_96-116},
\ref{tab:vv_128-136}, and \ref{tab:vv_150} list, respectively,
the occupation probabilities $v^2_j$ computed with the
zero-deformation constrained RHB and HFB calculations,
corresponding to the single-neutron and single-proton
configurations considered for those
odd-odd nuclei with masses $A=48-82$
($^{48}$Sc, $^{76}$As, and $^{82}$Br),
$A=96-116$ ($^{96}$Nb, $^{100}$Tc, and $^{116}$In)
$A=128-136$ ($^{128}$I, $^{130}$I, and $^{136}$Cs), and
$A=150$ ($^{150}$Pm).
These $v^2_j$ values are used to determine the
pair structure constants $\alpha_j$ \eqref{eq:alpha}
and $\beta_{j_1j_2}$ \eqref{eq:beta},
necessary ingredients to compute the coefficients
$A_{\rho}(j)$ and $B_{\rho}(j_1j_2)$,
which appear in the $\znbb$ operators \eqref{eq:mzn-1}.

\begin{table}
\caption{\label{tab:para-ddpc}
Strength parameters for the IBM-2
Hamiltonian (\ref{eq:hb}) obtained from 
the mapping procedure that is based on the RHB method.
The parameters of the Hamiltonian (\ref{eq:hb-semimagic})
adopted for $^{116}$Sn and $^{136}$Xe are as follows. 
$\epsilon_{d_{\nu}}=1.5$ MeV, $\kappa_\nu=-0.05$ MeV, 
and $\chi_{\nu}=0.80$ for $^{116}$Sn, 
and $\epsilon_{d_{\pi}}=1.5$ MeV, $\kappa_\pi=-0.04$ MeV, 
and $\chi_{\pi}=-0.80$ for $^{136}$Xe. 
}
 \begin{center}
 \begin{ruledtabular}
  \begin{tabular}{lccccc}
 & $\epsilon_d$ (MeV) & $\kappa$ (MeV) 
 & $\chi_{\nu}$ & $\chi_{\pi}$ & $\theta$ (MeV) \\
\hline
$^{48}$Ti & $0.65$ & $-0.70$ & $-1.30$ & $-1.30$ & $0.00$ \\ 
$^{76}$Ge & $0.60$ & $-0.38$ & $-0.90$ & $-0.50$ & $0.50$ \\ 
$^{76}$Se & $0.96$ & $-0.22$ & $0.90$ & $0.50$ & $0.30$ \\ 
$^{82}$Se & $0.80$ & $-0.70$ & $-1.00$ & $-1.00$ & $0.60$ \\ 
$^{82}$Kr & $1.16$ & $-0.35$ & $-0.40$ & $-0.40$ & $0.40$ \\ 
$^{96}$Zr & $0.93$ & $-0.35$ & $-0.45$ & $0.47$ & $-0.20$ \\ 
$^{96}$Mo & $0.75$ & $-0.44$ & $-0.65$ & $0.45$ & $0.00$ \\ 
$^{100}$Mo & $0.58$ & $-0.35$ & $-0.50$ & $0.45$ & $0.20$ \\ 
$^{100}$Ru & $0.50$ & $-0.44$ & $-0.70$ & $-0.40$ & $0.40$ \\ 
$^{116}$Cd & $0.85$ & $-0.23$ & $-0.30$ & $0.40$ & $0.00$ \\ 
$^{128}$Te & $0.78$ & $-0.48$ & $0.40$ & $-0.90$ & $0.00$ \\ 
$^{128}$Xe & $0.42$ & $-0.42$ & $0.40$ & $-0.80$ & $0.00$ \\ 
$^{130}$Te & $0.95$ & $-0.48$ & $0.30$ & $-0.78$ & $0.00$ \\ 
$^{130}$Xe & $0.56$ & $-0.44$ & $0.25$ & $-0.86$ & $0.40$ \\ 
$^{136}$Ba & $0.99$ & $-0.25$ & $-0.56$ & $-0.95$ & $0.00$ \\ 
$^{150}$Nd & $0.16$ & $-0.24$ & $-0.80$ & $-0.50$ & $0.00$ \\ 
$^{150}$Sm & $0.16$ & $-0.21$ & $-0.70$ & $-0.55$ & $0.00$ \\ 
  \end{tabular}
 \end{ruledtabular}
 \end{center}
\end{table}

\begin{table}
\caption{\label{tab:para-d1m}
Same as the caption to Table~\ref{tab:para-ddpc},
but the HFB method is used.
The parameters
$\epsilon_{d_{\nu}}=1.6$ MeV, $\kappa_\nu=-0.05$ MeV, 
and $\chi_{\nu}=0.70$ for $^{116}$Sn;
and $\epsilon_{d_{\pi}}=1.5$ MeV, $\kappa_\pi=-0.04$ MeV, 
and $\chi_{\pi}=-0.80$ for $^{136}$Xe. 
}
 \begin{center}
 \begin{ruledtabular}
  \begin{tabular}{lccccc}
 & $\epsilon_d$ (MeV) & $\kappa$ (MeV) 
 & $\chi_{\nu}$ & $\chi_{\pi}$ & $\theta$ (MeV) \\
\hline
$^{48}$Ti & $1.10$ & $-0.38$ & $-1.30$ & $-1.30$ & $0.00$ \\ 
$^{76}$Ge & $0.83$ & $-0.27$ & $-0.85$ & $-0.65$ & $0.50$ \\ 
$^{76}$Se & $1.00$ & $-0.14$ & $0.40$ & $0.40$ & $0.20$ \\ 
$^{82}$Se & $0.76$ & $-0.38$ & $-1.13$ & $-1.13$ & $0.50$ \\ 
$^{82}$Kr & $1.16$ & $-0.25$ & $-0.56$ & $-0.56$ & $0.40$ \\ 
$^{96}$Zr & $1.24$ & $-0.25$ & $-0.25$ & $0.47$ & $0.00$ \\ 
$^{96}$Mo & $1.00$ & $-0.16$ & $-0.65$ & $-0.65$ & $0.00$ \\ 
$^{100}$Mo & $0.58$ & $-0.16$ & $0.03$ & $0.10$ & $0.10$ \\ 
$^{100}$Ru & $1.18$ & $-0.29$ & $-1.00$ & $-1.00$ & $0.40$ \\ 
$^{116}$Cd & $0.84$ & $-0.26$ & $-0.72$ & $-0.47$ & $0.00$ \\ 
$^{128}$Te & $0.92$ & $-0.26$ & $0.40$ & $-0.90$ & $0.00$ \\ 
$^{128}$Xe & $0.60$ & $-0.30$ & $0.25$ & $-0.45$ & $0.30$ \\ 
$^{130}$Te & $1.08$ & $-0.28$ & $0.30$ & $-0.78$ & $0.00$ \\ 
$^{130}$Xe & $0.75$ & $-0.27$ & $0.25$ & $-0.86$ & $0.30$ \\ 
$^{136}$Ba & $1.05$ & $-0.23$ & $-1.10$ & $-1.10$ & $0.00$ \\ 
$^{150}$Nd & $0.13$ & $-0.23$ & $-0.70$ & $-0.50$ & $0.00$ \\ 
$^{150}$Sm & $0.16$ & $-0.17$ & $-0.60$ & $-0.55$ & $0.00$ \\
  \end{tabular}
 \end{ruledtabular}
 \end{center}
\end{table}

\subsection{Parameters for the IBM-2\label{sec:para}}

The derived IBM-2 strength parameters 
are given in Table~\ref{tab:para-ddpc} 
and Table~\ref{tab:para-d1m}, which are determined
by performing the RHB and HFB SCMF calculations
to provide microscopic inputs 
for the mapping procedure, respectively. 
Most of those parameters derived based on the RHB
are the same as those employed in the calculations
of the $\tnbb$-decay
NMEs in Refs.~\cite{nomura2022bb,nomura2024bb}, 
as shown in Table~IX of \cite{nomura2022bb} 
and Fig.~4 of \cite{nomura2024bb}. 
Here, for many of the nuclei the cubic term 
is included, which was not considered in 
\cite{nomura2022bb,nomura2024bb} due to 
a limitation of the computer code. 
There are also slight 
modifications that do not affect the final results. 
For instance, the $\hat L \cdot \hat L$ term 
is not included in the present IBM-2 Hamiltonian
either with the RHB or HFB input, 
while this term was introduced in a few nuclei in
Refs.~\cite{nomura2022bb,nomura2024bb}.
The negative $\theta$ value chosen for $^{96}$Zr 
in Table~\ref{tab:para-d1m} is to create 
a prolate and an oblate minima, as was done 
in Ref.~\cite{nomura2020zr}.

\acknowledgements
This work has been supported by JSPS
KAKENHI Grant No. JP25K07293.

\bibliography{refs}

\end{document}